\documentclass[letterpaper,oneside,final,notitlepage,onecolumn]{article}
\usepackage{graphicx}
\usepackage{amsmath}
\usepackage{amsfonts}
\usepackage{amssymb}

\begin{document}

\begin{center}
{\LARGE Hierarchic Theory of Complex Systems (biosystems, colloids):}

\smallskip

{\Large self-organization \& osmos}

\medskip

{\Large Alex Kaivarainen}

\medskip

{\large JBL, \ University of Turku, \ FIN-20520, Turku, Finland\medskip}

{\large URL: \thinspace http://www.karelia.ru/\symbol{126}alexk}

{\large H2o@karelia.ru}

\medskip

\medskip
\end{center}

\begin{quotation}
\thinspace\thinspace\textbf{Materials, presented in this original article are
based on following publications:}

\smallskip

\textbf{[1]. A. Kaivarainen. Book: Hierarchic Concept of Matter and Field.
Water, biosystems and elementary particles. New York, NY, 1995 and new version
of this book. }

\textbf{(see} \textbf{URL: http://www.karelia.ru/\symbol{126}alexk [Book
prospect and New articles]).}

\textbf{[2]. \thinspace A. Kaivarainen. New Hierarchic Theory of Matter
General for Liquids and Solids: }

\textbf{dynamics, thermodynamics and mesoscopic structure of water and ice }

\textbf{(see URL: http://www.karelia.ru/\symbol{126}alexk [New articles]).}

\textbf{[3].} \textbf{A. Kaivarainen. Hierarchic Concept of Condensed Matter
and its Interaction with Light: New Theories of Light Refraction, Brillouin
Scattering\ and M\"{o}ssbauer effect }

\textbf{(see URL: http://www.karelia.ru/\symbol{126}alexk \ [New articles]). }

\textbf{[4]. A. Kaivarainen. Hierarchic Theory of Condensed Matter:
Interrelation between mesoscopic and macroscopic properties (see URL:
http://www.karelia.ru/\symbol{126}alexk \ \ [New articles]). }

\smallskip

\textbf{Computerized verification of described here new theories is presented,
using special computer program, based on Hierarchic Theory of Condensed Matter
(copyright, 1997, A. Kaivarainen).}

\smallskip

\textbf{CONTENTS OF\ ARTICLE}

\textbf{\medskip}
\end{quotation}

\begin{quote}
\textbf{Summary to Part I of book: ''Hierarchic theory of condensed matter''}

\textbf{Introduction}
\end{quote}

\textbf{1.\ Protein domain mesoscopic organization}

\textbf{2. Quantum background of lipid domain organization in biomembranes}

\textbf{3.\ Hierarchic approach to theory of solutions and colloid systems}

\textbf{4. Distant solvent-mediated interaction between macromolecules}

\textbf{5. Spatial self-organization in the water-macromolecular systems}

\textbf{6. Properties of [bisolvent - polymer system]}

\textbf{7. Osmosis and solvent activity. Traditional and mesoscopic approach}

\bigskip

\begin{center}
=====================================================================\bigskip

{\large Summary of}

{\large New Hierarchic Theory of Condensed Matter (see
http://arXiv.org/abs/physics/0003044) }
\end{center}

{\large \smallskip}

\textbf{\ A basically new hierarchic quantitative theory, general for solids
and liquids, has been developed.}

\textbf{It is assumed, that anharmonic oscillations of particles in any
condensed matter lead to emergence of three-dimensional (3D) superposition of
standing de Broglie waves of molecules, electromagnetic and acoustic waves.
Consequently, any condensed matter could be considered as a gas of 3D standing
waves of corresponding nature. Our approach unifies and develops strongly the
Einstein's and Debye's models. }\ \textbf{Collective excitations, like 3D
standing de Broglie waves of molecules, representing at certain conditions the
mesoscopic molecular Bose condensate, were analyzed, as a background of
hierarchic model of condensed matter.}

\smallskip

\textbf{The most probable de Broglie wave (wave B) length is determined by the
ratio of Plank constant to the most probable impulse of molecules, or by ratio
of its most probable phase velocity to frequency. The waves B are related to
molecular translations (tr) and librations (lb).}

As the quantum dynamics of condensed matter does not follow in general case
the classical Maxwell-Boltzmann distribution, the real most probable de
Broglie wave length can exceed the classical thermal de Broglie wave length
and the distance between centers of molecules many times.

\textit{This makes possible the atomic and molecular Bose condensation in
solids and liquids at temperatures, below boiling point. It is one of the most
important results of new theory, which we have confirmed by computer
simulations on examples of water and ice.}

\smallskip

\textbf{Four strongly interrelated }new types of quasiparticles (collective
excitations) were introduced in our hierarchic model:

1.~\textit{Effectons (tr and lb)}, existing in "acoustic" (a) and "optic" (b)
states represent the coherent clusters in general case\textbf{; }

2.~\textit{Convertons}, corresponding to interconversions between \textit{tr
}and \textit{lb }types of the effectons (flickering clusters);

3.~\textit{Transitons} are the intermediate $\left[  a\rightleftharpoons
b\right]  $ transition states of the \textit{tr} and \textit{lb} effectons;

4.~\textit{Deformons} are the 3D superposition of IR electromagnetic or
acoustic waves, activated by \textit{transitons }and \textit{convertons. }\smallskip

\smallskip

\ \textbf{Primary effectons }(\textit{tr and lb) }are formed by 3D
superposition of the \textbf{most probable standing de Broglie waves }of the
oscillating ions, atoms or molecules. The volume of effectons (tr and lb) may
contain from less than one, to tens and even thousands of molecules. The first
condition means validity of \textbf{classical }approximation in description of
the subsystems of the effectons. The second one points to \textbf{quantum
properties} \textbf{of coherent clusters due to molecular Bose condensation}%
\textit{. }

\ The liquids are semiclassical systems because their primary (tr) effectons
contain less than one molecule and primary (lb) effectons - more than one
molecule. \textit{The solids are quantum systems totally because both kind of
their primary effectons (tr and lb) are molecular Bose condensates.}%
\textbf{\ These consequences of our theory are confirmed by computer
calculations. }

\ The 1st order $\left[  gas\rightarrow\,liquid\right]  $ transition is
accompanied by strong decreasing of rotational (librational) degrees of
freedom due to emergence of primary (lb) effectons and $\left[
liquid\rightarrow\,solid\right]  $ transition - by decreasing of translational
degrees of freedom due to Bose-condensation of primary (tr) effectons.

\ \textbf{In the general case the effecton can be approximated by
parallelepiped with edges corresponding to de Broglie waves length in three
selected directions (1, 2, 3), related to the symmetry of the molecular
dynamics. In the case of isotropic molecular motion the effectons' shape may
be approximated by cube.}

\textbf{The edge-length of primary effectons (tr and lb) can be considered as
the ''parameter of order''.}

\smallskip

The in-phase oscillations of molecules in the effectons correspond to the
effecton's (a) - \textit{acoustic }state and the counterphase oscillations
correspond to their (b) - \textit{optic }state. States (a) and (b) of the
effectons differ in potential energy only, however, their kinetic energies,
impulses and spatial dimensions - are the same. The \textit{b}-state of the
effectons has a common feature with \textbf{Fr\"{o}lich's polar mode. }

\smallskip

\textbf{The }$(a\rightarrow b)$\textbf{\ or }$(b\rightarrow a)$%
\textbf{\ transition states of the primary effectons (tr and lb), defined
as\ primary transitons, are accompanied by a change in molecule polarizability
and dipole moment without density fluctuations. At this case they lead to
absorption or radiation of IR photons, respectively.}

\textbf{\ Superposition (interception) of three internal standing IR photons
of different directions (1,2,3) - forms primary electromagnetic deformons (tr
and lb).}

\ On the other hand, the [lb$\rightleftharpoons\,$tr] \textit{convertons }and
\textit{secondary transitons} are accompanied by the density fluctuations,
leading to \textit{absorption or radiation of phonons}.

\textit{Superposition resulting from interception} of standing phonons in
three directions (1,2,3), forms \textbf{secondary acoustic deformons (tr and
lb). }

\smallskip

\ \textit{Correlated collective excitations }of primary and secondary
effectons and deformons (tr and lb)\textbf{, }localized in the volume of
primary \textit{tr }and \textit{lb electromagnetic }deformons\textbf{, }lead
to origination of \textbf{macroeffectons, macrotransitons}\textit{\ }and
\textbf{macrodeformons }(tr and lb respectively)\textbf{. }

\ \textit{Correlated simultaneous excitations of \thinspace tr and lb}
\textit{macroeffectons }in the volume of superimposed \textit{tr }and
\textit{lb }electromagnetic deformons lead to origination of
\textbf{supereffectons. }

\ In turn, the coherent excitation of \textit{both: tr }and \textit{lb
macrodeformons and macroconvertons }in the same volume means creation of
\textbf{superdeformons.} Superdeformons are the biggest (cavitational)
fluctuations, leading to microbubbles in liquids and to local defects in solids.

\smallskip

\ \textbf{Total number of quasiparticles of condensed matter equal to 4!=24,
reflects all of possible combinations of the four basic ones [1-4], introduced
above. This set of collective excitations in the form of ''gas'' of 3D
standing waves of three types: de Broglie, acoustic and electromagnetic - is
shown to be able to explain virtually all the properties of all condensed matter.}

\ \textit{The important positive feature of our hierarchic model of matter is
that it does not need the semi-empiric intermolecular potentials for
calculations, which are unavoidable in existing theories of many body systems.
The potential energy of intermolecular interaction is involved indirectly in
dimensions and stability of quasiparticles, introduced in our model.}

{\large \ The main formulae of theory are the same for liquids and solids and
include following experimental parameters, which take into account their
different properties:}

$\left[  1\right]  $\textbf{- Positions of (tr) and (lb) bands in oscillatory spectra;}

$\left[  2\right]  $\textbf{- Sound velocity; }$\,$

$\left[  3\right]  $\textbf{- Density; }

$\left[  4\right]  $\textbf{- Refraction index (extrapolated to the infinitive
wave length of photon}$)$\textbf{.}

\textit{\ The knowledge of these four basic parameters at the same temperature
and pressure makes it possible using our computer program, to evaluate more
than 300 important characteristics of any condensed matter. Among them are
such as: total internal energy, kinetic and potential energies, heat-capacity
and thermal conductivity, surface tension, vapor pressure, viscosity,
coefficient of self-diffusion, osmotic pressure, solvent activity, etc. Most
of calculated parameters are hidden, i.e. inaccessible to direct experimental measurement.}

\ The new interpretation and evaluation of Brillouin light scattering and
M\"{o}ssbauer effect parameters may also be done on the basis of hierarchic
theory. Mesoscopic scenarios of turbulence, superconductivity and superfluity
are elaborated.

\ Some original aspects of water in organization and large-scale dynamics of
biosystems - such as proteins, DNA, microtubules, membranes and regulative
role of water in cytoplasm, cancer development, quantum neurodynamics, etc.
have been analyzed in the framework of Hierarchic theory.

\medskip

\textbf{Computerized verification of our Hierarchic concept of matter on
examples of water and ice is performed, using special computer program:
Comprehensive Analyzer of Matter Properties (CAMP, copyright, 1997,
Kaivarainen). The new opto-acoustical device (CAMP), based on this program,
with possibilities much wider, than that of IR, Raman and Brillouin
spectrometers, has been proposed (see URL:\thinspace\thinspace
http://www.karelia.ru/\symbol{126}alexk).}

\smallskip

\textbf{This is the first theory able to predict all known experimental
temperature anomalies for water and ice. The conformity between theory and
experiment is very good even without any adjustable parameters. }

\textbf{The hierarchic concept creates a bridge between micro- and macro-
phenomena, dynamics and thermodynamics, liquids and solids in terms of quantum
physics. }\newpage

\begin{center}
{\Large Introduction}\bigskip
\end{center}

\textbf{Domain granular structure is pertinent to solid bodies, liquid
crystals, and polymers of artificial and biological origin. In liquids, as is
seen from the X-ray data, the local order is also kept like in solid bodies.
Just like in the case of solids, local order in liquids can be caused by the
most probable - primary effectons (see Introduction to [1] and [2]), but
smaller in size.}

In water, the relatively stable clusters of molecules are revealed by the
quasielastic neutron scattering method. The diameter of these clusters are
(20-30) \AA\ and the lifetime is of the order of $10^{-10}s$ $($Gordeyev and
Khaidarov, 1983). These parameters are close to those we have calculated for
librational water effectons (Fig. 7a of [1] or Fig4a of [2]).

A coherent-inelastic neutron scattering, performed on liquid $D_{2}O$ at room
temperature revealed collective high-frequency sound mode. Observed collective
excitation has a solid-like character with dimension around 20\AA, resembling
water clusters with saturated hydrogen bonds. The observed sound velocity in
these clusters is about 3300 m/s, i.e. close to velocity of sound in ice and
much bigger than that in liquid water: 1500 m/s (Teixeira et al., 1985). Such
data confirm the existence of primary librational effectons as molecular Bose
condensate, leading from our hierarchic theory.

Among the earlier theoretical models of water the model of ''flickering
clusters'' proposed by Frank and Wen (1957) is closer to our model than
others. The ''flickering'' of a cluster consisting of water molecules is
expressed by the fact that it dissociates and associates with a short period
$(10^{-10}-10^{-11})$ s. Near the non-polar molecules this period grows up
(Frank and Wen, 1957, Frank and Evans, 1945) and ''icebergs'' appear. The
formation of hydrogen bonds in water is treated as a cooperative process. Our
$[lb/tr]$ convertons, i.e. interconversions between primary lb and tr
effectons, reflect the properties of flickering clusters better than other
quasiparticles of hierarchic model.

Proceeding from the flickering cluster model, Nemethy and Scheraga (1962),
using the methods of statistical thermodynamics, calculated a number of
parameters for water (free energy, internal energy, entropy) and their
temperature dependences, which agree with the experimental data in the limits
of 3\%. However, calculations of heat capacity were less successful. The
quantity of water molecules decreases from 91 at $0^{0}C$ to 25 at $70^{0}C$
$($Nemethy and Scheraga, 1962). It is in rather good agreement with our
results (Fig. 7a of [1] or Fig.4a of [2]) on the change of the number of water
molecules in a primary librational effecton with temperature.

The stability of primary effectons (clusters, domains), forming the condensed
media, is determined by the coherence of heat motions, the equality of the
most probable 3D standing waves B of atoms (molecules), increasing distant Van
der Waals interaction in the volume of the effectons.

It could be possible that molecules, atoms, or ions of \textit{different
molecular masses }belong to the same effectons. The equality of wave B lengths
for such different particles, forming the effectons:%

\begin{equation}
\lambda_{1}={\frac{h}{m_{1}v_{1}}}=\lambda_{2}={\frac{h}{m_{2}v_{2}}%
}=...=\lambda_{i}={\frac{h}{m_{i}v_{i}}}\tag{1}%
\end{equation}
means that differences in masses are compensated by differences in the group
velocities of these particles so that their most probable impulses are equal:%

\begin{equation}
P_{i}=m_{1}v_{1}=m_{2}v_{2}=...=m_{i}v_{i}\tag{2}%
\end{equation}
\textbf{The domains or the crystallites in solid bodies, which could be
considered as a primary effectons, can contain a big number of elementary
cells. Transitions between the different types of elementary cells (second
order phase transitions) means cooperative redistribution in the positions and
dynamics of atoms, leading to origination of new primary effectons. In
accordance with our model, the second order phase transitions are related
sometimes also with emergency of conditions for primary effectons
polymerization or distant association in coherent superclusters, and
concomitant shifting of their }$\left(  a\Leftrightarrow b\right)
$\textbf{\ equilibrium to the left.}

\textbf{The mesoscopic theory could be used to describe a wide range of
physico-chemical and biological phenomena.}

\vspace{0in}

\begin{center}
{\large 1. Protein domain mesoscopic organization}
\end{center}

\smallskip

\textbf{If the geometry of cavities of protein surface are complementary to
the geometry of water librational effectons, then the latter are stabilized.
In the opposite case, the water effectons in cavities are either not realized,
or unstable. In that case, the probability of the water cluster dissociation
in the cavity, related to }$[lb\rightarrow tr]$\textbf{\ conversion,
increases. The evolution of biological macromolecules could have gone in such
a way that they ''learned'' to use the cooperative properties of water
clusters and their dissociation for regulation of their large-scale dynamics
and signal transmission. Calculated frequency of }$[lb/tr]$%
\textbf{\ convertons }$(10^{6}-10^{7})\,c^{-1}$\textbf{\ coincide with
frequency of protein cavities large-scale pulsations, accompanied by domains
relative fluctuations}.

All sufficiently large globular proteins consist of domains whose dimensions
under normal conditions vary in the narrow limits: (10-20)\AA\ (K\"{a}%
iv\"{a}r\"{a}inen, 1985). This value is close to dimensions of librational
water effecton (Fig. 7a of [1] or Fig. 4a of [2]) that confirms the important
role of water in evolution of biopolymers. We may predict that the lower is
the physiological temperature of given organism the larger are the interdomain
cavities and domains of its proteins. It is known that the water in pours or
cavities with diameter less than 50$\,$\AA$\;$freeze out at very low
temperatures (about $-60^{0}C)$ and its viscosity is high (Martini et al.,
1983). For the other hand, our calculations shows (Fig.17b of [1]), that
freezing in normal conditions should be accompanied by increasing the linear
size of primary librational effectons just till 50\ \AA.

\textbf{As far the condition for librational effectons growth in narrow pours
is absent, the formation of sufficiently big primary translational effectons
is also violated. As a result of that, condition (6.6 of [1]) for [liquid
}$\rightarrow$\textbf{\ solid] phase transition occurs in such cavities at
much lower temperature than in bulk water.}

If the sound velocity in proteins and the positions of maxima in their
oscillatory spectra, which characterize the librations of atoms \textit{are
known}, then the most probable wave B length of aminoacids groups and atoms,
forming the domains $(\lambda_{1},\lambda_{2},\lambda_{3})$ can be estimated.
If the volume of an effecton is approximated by a sphere:%

\begin{equation}
V_{ef}={\frac{9(\lambda_{1}\lambda_{2}\lambda_{3})}{4\pi}}={\frac{4}{3}}\pi
r^{3},\tag{3}%
\end{equation}
then its radius:%

\begin{equation}
r=\left[  {\frac{27(\lambda_{1}\lambda_{2}\lambda_{3})}{16\pi^{2}}}\right]
^{1/3}=0.555\lambda_{\text{res}},\tag{4}%
\end{equation}%
\[
\text{where: \ }\lambda_{\text{res}}=(\lambda_{1}\lambda_{2}\lambda_{3}%
)^{1/3}\text{,\ \thinspace and the diameter:\ \ \ }d=2r=1.11\lambda
_{\text{res}}.
\]
The collective properties of protein's primary effectons presented by $\alpha
$-structures, $\beta$-sheets and whole domains can determine the cooperative
properties of biopolymers.

\textbf{Heat oscillations of atoms and atom groups, forming the protein
effectons must be coherent, like in any other condensed matter.}

Such ideas agree with the Fr\"ochlich hypothesis about the possibility of
Bose-condensation in biological systems (Fr\"ohlich, 1975).

The notion of \ ''knots'' in proteins was introduced by R.Lumry and B. Gregory
(1986). The knots are regions, containing very slow $H\Leftrightarrow D$
exchangeable protons in composition of compact cooperative structures.

The dimensions of knots are less than dimensions of domains.

It looks that knots could represent a translational effectons, as far their
$(a\Leftrightarrow b)_{tr}\,$transitions, in contrast to $(a\Leftrightarrow
b)_{lb}$ ones of librational effectons do not accompanied by reorganization of
hydrogen bonds. Consequently, the possibility for $H\Leftrightarrow D$
exchange in knots is more limited.

The molecular dynamic computer simulations of proteins reveal, indeed, a
highly correlated collective motion of groups of atoms, inhomogeneously
distributed in proteins structure (Swaminathan et al., 1982).

It looks that the traditional theory of protein tertiary native structure
self-organization from primary one (Cantor and Shimmel, 1980) is not totally
successful as far it does not take into account quantum process, related to
formation the protein effectons as 3D standing waves B of protein atoms.

The change of interdomain interactions, the stabilization of its small-scale
dynamics of proteins by ligands leading to the increase of $\lambda
_{\text{res}}$ (see eq.4) as a measure of cooperativeness, can provide the
long- distance signal transmission in macromolecules and allosteric effects in
oligomeric proteins. The $[lb/tr]$ convertons, i.e. dissociation of
librational water effectons in the protein cavities is the key phenomenon in
the above mentioned processes.

The described events interrelate the small-scale dynamics of atoms and the
large-scale dynamics of domains and subunits to the dynamics of water clusters
in protein interdomain cavities, dependent in turn on the properties of bulk
water (K\"aiv\"ar\"ainen, 1985,1989b, K\"aiv\"ar\"ainen et al., 1993).

Our mesoscopic mechanism of signal transmission in proteins is alternative to
\textit{Davidov}'\textit{s soliton }mechanism. The latter is good only for
highly ordered systems with small dissipation..

\medskip

\begin{center}
{\large 2. Quantum background of lipid domain organization in biomembranes}
\end{center}

\smallskip

The importance of lipid domains in membranes for their functioning is known,
but the physical background for domains origination remains unclear.

Our mesoscopic theory was used for computer simulations of lipid domain
dimensions in model membranes. The known data on the position of IR bands,
corresponding to asymmetric $\left[  N-(CH_{3})_{3}\right]  $ stretching in
holine for \textit{trance} $(920cm^{-1})$ and \textit{gauche} (900 and
$860cm^{-1}) $ conformations where taken for calculations. The knowledge of
sound velocity and its changes as a result of phase transitions:
$(1.97\cdot10^{5}cm/s)^{38^{o}}\rightarrow(1.82\cdot10^{5}cm/s)^{42^{o}}$ is
also necessary for calculations of the domain dimensions, using eq.(2.59 of
[1]). The results of calculations are presented on Fig.1. In our approach the
lipid domains in biomembranes and their artificial models are considered as
quasiparticles-primary librational -\textit{effectons}, formed by 3D
superposition of the most probable de Broglie waves $(\lambda=h/p)_{1,2,3}$,
determined by coherent thermal oscillations of lipid molecules. The lesser the
value of the most probable impulse $(p=mv)_{1,2,3}$ of lipid, the bigger is
corresponding $\lambda_{1,2,3}$ and the effecton volume (eq.3).

According to our calculations, a rise in temperature from 0 to $70^{0}$ leads
to a decrease in the most probable $\lambda$ from 88 to 25 \AA. In the phase
transition region $(38-42^{0})\;\lambda$ decreases from 46 to 37 \AA
\thinspace(Fig.1). The former process corresponds to change of the lipid
domains volume from (50 to $2)\cdot10^{4}\AA^{3}$ and the latter one from (7.5
to $5)\cdot10^{4}\AA^{3}$, respectively (Fig. 2). The values of these changes
coincide with available experimental data. Like the calculations we made
earlier for ice and water, these results provide further support of our
theoretical approach.

\begin{center}%
\begin{center}
\includegraphics[
height=2.0029in,
width=4.9692in
]%
{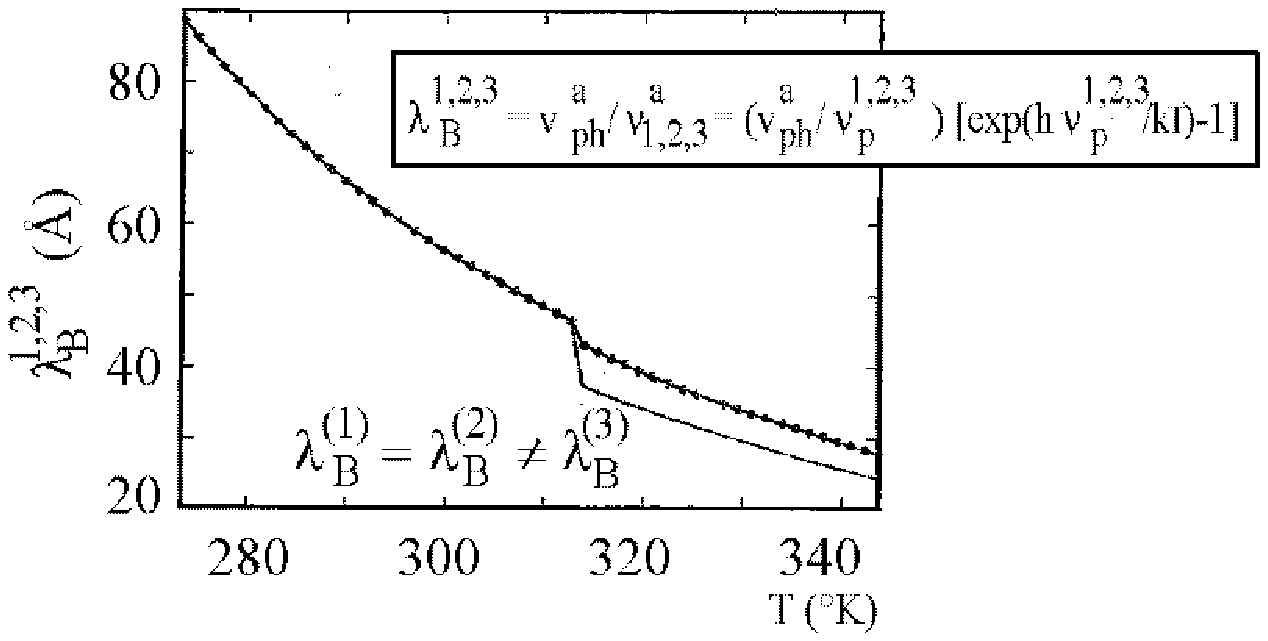}%
\end{center}
\end{center}

\begin{quotation}
\textbf{Fig.~1. }Temperature dependencies of the most probable de Broglie wave
length of lipids $(\lambda_{1},\lambda_{2},\lambda_{3})$, related to their
\textit{stretching}. The values of $\lambda_{1},\lambda_{2},\lambda_{3} $
determine the spatial dimensions of lipid domains. Domains are considered as
quasiparticles (primary effectons), formed by 3D superposition of the most
probable de Broglie waves of lipids.\medskip
\end{quotation}

\begin{center}%
\begin{center}
\includegraphics[
height=2.1923in,
width=4.9528in
]%
{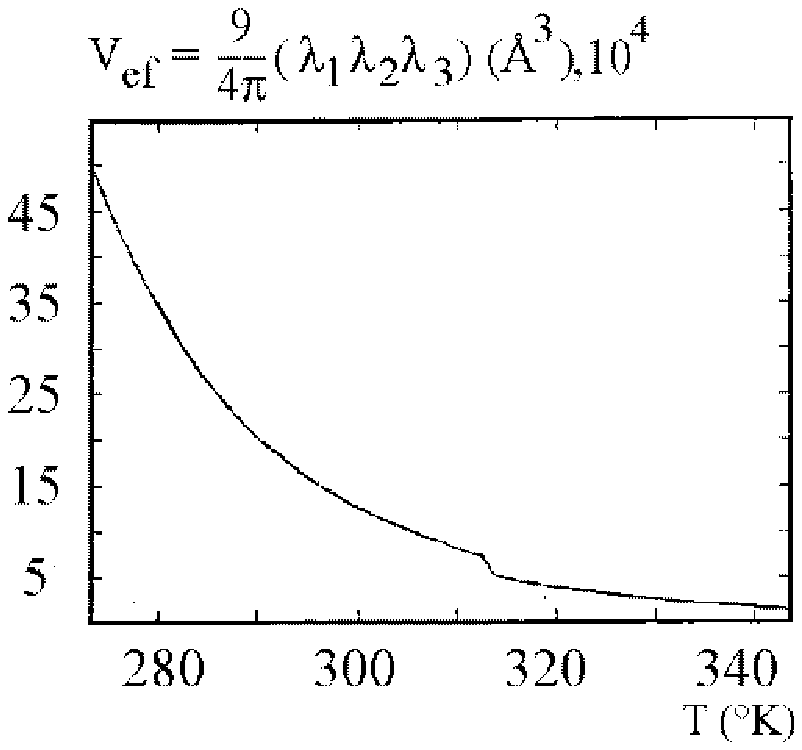}%
\end{center}
\medskip
\end{center}

\begin{quotation}
\textbf{Fig.~2. }Temperature dependence of the volume of lipid domains:
$V={\frac{9}{4\pi}}\lambda_{1}\lambda_{2}\lambda_{3}$.

The volume is determined by 3D superposition of the most probable de Broglie
waves of lipid molecules or their fragments. Phase transition occurs in the
region of $312K\;(39^{0})$ and accompanies by \textit{trance - gauche} change
of lipid conformation.
\end{quotation}

\medskip

\textbf{As discussed in Chapter 17 of [1] and my paper: Hierarchic Model of
Consciousness in URL:\ [http://kftt.karelia.ru/\symbol{126}alexk/papers] the
mesoscopic organization of biomembranes, related to their dynamics and that of
microtubules system - may play an important role in elementary act of
consciousness. }

\medskip

\begin{center}
{\large 3. Mesoscopic approach to theory of solutions}
\end{center}

{\large \smallskip}

\textbf{The action of the dissolved molecules can lead to the shift of
}$\left(  a\Leftrightarrow b\right)  $\textbf{\ equilibrium to the right or to
the left for effectons (tr and lb) of solvent. In the former case the lifetime
of unstable state for primary effectons increases, and in the latter case the
stabilization of molecular associates (clusters) takes place.}

\textbf{The same is true for convertons equilibrium: }$[lb\Leftrightarrow
tr]$\textbf{, reflecting [association }$\Leftrightarrow$%
\textbf{\ dissociation] of water clusters (primary librational effectons).}

\textbf{The effects of stabilization of clusters can be reinforced at such
concentration of dissolved molecules, when the mean distances between them (r)
coincide with one of the primary effectons ribs }$(\lambda_{1},\lambda
_{2},\lambda_{3})$\textbf{\ or their integer number:}%

\begin{equation}
r={\frac{11.8}{(C_{M})^{1/3}}}(\AA)\tag{5}%
\end{equation}
where C$_{M}$ is the molar concentration.

\smallskip

So the concentration dependence of the stabilizing action brought by the
dissolved substance upon the solvent can be non-monotonic and periodic. Such
data have been published indeed (Tereshkevitch et al., 1974).

If the wave B length of the dissolved molecule or atoms exceeds the dimensions
of primary effectons, then it must increase the degree of liquid association.
In the opposite case the ordering of liquid structure decreases. To prove the
aforementioned, it should be noted that the \textit{structure-forming }ions,
with a positive hydration $(Li^{+},\,\,Na^{+},\,\,F^{-})$, as a rule, have a
lesser mass and lesser impulse (i.e. a larger value of $\lambda_{B}%
=h/mv_{gr})$ than those with a negative hydration $(Rb^{+},\;Cs^{+}%
,\;Br^{-},\,\,I^{-})$ with the same charges.

In accordance with such ideas, nonpolar atoms, minimally distorting the
properties of a pure liquid, e.g. He, Ne, Ar, have a maximal structuring
action, stabilizing \textit{a}-states of primary librational effectons.

{\large Our Hierarchic model may be useful for elaboration a general theory of solutions.}

\textbf{In host-guest systems a following situations are possible: }

\textbf{1) guest molecules stabilize (a)-states of host effectons (tr and lb)
and increases their dimensions. The }$(a\Leftrightarrow b)$%
\textbf{\ equilibrium of the effectons and }$[lb\Leftrightarrow tr]$%
\textbf{\ equilibrium of convertons becomes shifted leftward decreasing
potential energy of a system, corresponding to its stabilization effect;}

\textbf{2) guest molecules destabilize (a)-states of host effectons. The
}$(a\Leftrightarrow b)$\textbf{\ and }$[lb\Leftrightarrow tr]$%
\textbf{\ equilibriums of the primary effectons and convertons correspondingly
are shifted rightward, inducing general destabilization effect of the system;}

\textbf{3) guest and host molecules form separate individual effectons
(mesophase) without separation in two macrophases;}

\textbf{\smallskip}

With the increasing concentration of guest in solution of two molecular
liquids (for example, water - ethanol) the roles of guest and host may change.

It was shown that conductivity of aqueous solutions of NaCl, containing ions:
$Na^{+},\,Cl^{-},\,H_{3}O^{+}$ and OH$^{-}$ show to vary in a different linear
fashion over two ranges of temperature: $273\le T\le323K$ and $323\le T\le
360$K. The change in slope of the plot shows transition in the character of
water-ions interaction near 323 K (Roberts et al.1994). In the same work was
revealed, that the above aqueous systems exhibited some ''memory'' of the
temperature effects after changing the temperature from low to high and then
from high to low values. Such memory could mean a slow relaxation process,
accompanied by redistribution between populations of different excitations and
their equilibrium constants, i.e. the new type of self-organization in solution.

\textbf{In accordance with our theory, all reorganizations of liquid's
properties must be accompanied by correlated changes in the following
parameters of solutions:}

\textbf{1) sound velocity;}

\textbf{2) positions of translational and librational bands in oscillatory spectra;}

\textbf{3) density;}

\textbf{4) refraction index;}

\textbf{5) share and bulk viscosities;}

\textbf{6) coefficient of self-diffusion}

\textbf{7) light scattering;}

\textbf{8) heat capacity and thermal conductivity;}

\textbf{9) vapor pressure and surface tension.}

\textbf{\smallskip}

There are also a lot of other parameters and properties that should change in
solutions as a result of solute-solvent interaction. It could be revealed by
computer simulations, using software, elaborated.

The first four of the above listed parameters are present in the main formulas
of Hierarchic theory and must be determined under similar conditions
(temperature, pressure, etc.).

There are some experimental data which are in general agreement with the
consequences of our theory, pointing to interrelation between the above listed
parameters. The changes of sound velocity in different water - ethanol
mixtures as well as that of light and neutron scattering were studied in
detail by D'Arrigo and Paparelli $(1988a,1988b,1989)$, Benassi et al. (1988),
D'Arrigo and Teixeira (1990).

Correlations between density, viscosity, the refractive index, and the
dielectric constant of mixed molecular liquids at different temperatures were
investigated by D'Aprano's group in Rome (D'Aprano et al., $1989,1990a,1990b)$.

The interaction of a solute (guest) molecule with \textit{librational }solvent
effectons can be subdivided into \textit{two cases}: when the rotational
correlation time of a guest molecule $(\tau_{M}^{\text{rot}})$ is \textit{less
(a)} and \textit{more (b)} than the rotational correlation time of
\textit{librational effectons} $(\tau_{ef}^{lb})$:%

\begin{equation}
a)\qquad\tau_{M}^{\text{rot}}<\tau_{ef}^{lb}\text{ }\tag{6}%
\end{equation}
and%

\begin{equation}
b)\qquad\tau_{M}^{\text{rot}}>\tau_{ef}^{lb}\text{ }\tag{7}%
\end{equation}
In accordance with the Stokes-Einstein formula the corresponding rotational
correlation times:%

\begin{equation}
\tau_{M}^{\text{rot}}={\frac{V_{M}}{k}}{\frac{\eta}{T}\ \;\;}\text{and\ \ \ }%
\tau_{ef}^{lb}={\frac{V_{ef}^{lb}}{k}}\cdot{\frac{\eta}{T}}\tag{8}%
\end{equation}

where: $\eta$ and T are the share viscosity and absolute temperature of the solvent;

$\tau_{M}^{\text{rot}}$ and $\tau_{ef}^{lb}$ are dependent on the effective
volumes of a guest molecule $(V_{M})$ and the volume of primary librational effecton:%

\begin{equation}
V_{ef}^{lb}=n_{M}^{lb}(V_{0}/N_{0})={\frac{9}{4}}\pi(\lambda_{1}\lambda
_{2}\lambda_{3})\tag{9}%
\end{equation}
$\lambda_{1,2,3}$ are most probable wave B length in 3 selected directions;
$n_{M}^{\text{lib}}$ - number of molecules in a librational primary effecton,
depending on temperature: in water it decreases from 280 till to 3 in the
temperature interval $0-100^{0}C$ $($Fig. 7a of [1] or Fig. 4a of [2]).

When the condition (6) is realized, small and neutral guest molecules affect
presumably only the translational effectons.

In the second case (7) guest macromolecules can change the properties of both
types of effectons: translational and librational and shift the equilibrium
$[lb\Leftrightarrow tr]$ of convertons to the left, stimulating the cluster-formation.

In accordance with our model, \textbf{hydrophilic interaction }is related to
the shift of the $(a\Leftrightarrow b)$ equilibrium of \textit{translational}
effectons to the \textit{left}. As far the potential energy of the
(\textit{a}) state $(V_{a})$ is less than that in the (\textit{b}) state
$(V_{b})$, it means that such solvent-solute (host- guest) interaction will
decrease the potential and free energy of the solution. Hydrophilic
interaction does not need the realization of condition (7).

\textbf{Hydrophobic interaction}\textit{\ }is related to the shift of the
$(a\Leftrightarrow b)_{lb}$ equilibrium of \textit{librational effectons} to
the \textit{right}. Such a shift results in the increased potential energy of
the system. The dimensions of coherent water clusters representing librational
effectons under condition (7) may even increase. However, the decrease in of
entropy $(\Delta S)$ in this case is more than that in enthalpy $(\Delta H)$
and, consequently, free energy will increase: $\Delta G=\Delta H-T\Delta S>0$.
This is a source of hydrophobic interaction, leading to aggregation of
hydrophobic particles.

\bigskip

\textbf{Clusterphilic interaction}\textit{\ }was introduced by the author in
1980 already (K\"{a}iv\"{a}r\"{a}inen, 1980, 1985) to describe the cooperative
water cluster interaction with nonpolar protein cavities. This idea has got
support in the framework of our hierarchic concept. \textbf{Clusterphilic
interaction is related to the leftward shift of $(a\Leftrightarrow b)_{lb}$
equilibrium of primary librational effectons under condition (7), similar
shift of the equilibrium of $[lb\Leftrightarrow tr]$ of convertons and
increasing of lb effectons dimensions due to water immobilization (eqs.. 1 and 9)}

The latter effect is a result of decreasing of the rotational correlation time
of librational effectons and decreasing of the most probable impulses of water
molecules (2), related to librations under the effect of guest particles.
\textbf{Clusterphilic interactions can be subdivided into:}

\textbf{1}.\textbf{Intramolecular}\textit{\ -when water cluster is placed }in
the ''open'' states of big interdomain or intersubunit cavities and

\textbf{2. Intermolecular clusterphilic interaction}\textit{. Intermolecular
clusterphilic interactions }can be induced by very different sufficiently big macromolecules.

\smallskip

Clusterphilic interactions can play an important role in the self-organization
of biosystems, especially multiglobular allosteric enzymes, microtubules and
the actin filaments. \textbf{Cooperative properties of the cytoplasm,
formation of thixotropic structures, signal transmission in biopolymers,
membranes and distant interactions between different macromolecules can be
mediated by both types of clusterphilic interaction}s.

From (4.4 of [1, 2]) the contributions of primary translational and
librational effectons to total internal energy are:%

\begin{equation}
U_{ef}^{tr,lb}={\frac{V_{0}}{Z}}\left[  n_{ef}\left(  P_{ef}^{a}E_{ef}%
^{a}+P_{ef}^{b}E_{ef}^{b}\right)  \right]  _{tr,lb}\tag{10}%
\end{equation}
The contributions of this type of effectons to total kinetic energy (see 4.33) are:%

\begin{equation}
T_{ef}^{tr,lb}={\frac{V_{0}}{Z}}\left[  n_{ef}{\frac{\sum_{1}^{3}%
(E^{a})_{1,2,3}^{2}}{2m(V_{ph}^{a})^{2}}}\left(  P_{ef}^{a}+P_{ef}^{b}\right)
\right]  _{tr,lb}\tag{10a}%
\end{equation}
Subtracting (10a) from (10), we get the potential energy of primary effectons:%

\begin{equation}
V_{ef}=V_{tr}+V_{lb}=(U_{ef}-T_{ef})_{tr}+(U_{ef}-T_{ef})_{lb}\tag{11}%
\end{equation}

\textit{Clusterphilic interaction }and possible self-organization is promoted
mainly by \textit{decreasing }of V$_{lb}$ in the presence of macromolecules.

\textit{Hydrophilic }interaction, in accordance with our model, is a result of
$V_{tr}$. decreasing.

On the other hand, \textit{hydrophobic interaction }is a consequence of
$V_{lb}$ and $V_{tr}$ increasing in the presence of guest molecules.

\textit{Clusterphilic interaction }has been revealed, for example, in
dependencies of freezing temperature $(T_{f})$ for buffer solutions of
polyethyleneglycol (PEG) on its molecular mass and concentration
(K\"{a}iv\"{a}r\"{a}inen, 1985). The anomalous increasing of $T_{f}$ in the
presence of PEG with molecular mass more than 2000 D and at concentration less
than $30mg/ml$, pointing to increasing water activity, were registered by the
cryoscopy method. It may be explained as a result of clusterphilic interaction
increasing, when the fraction of ice-like water structures with saturated
hydrogen bonds, presented by primary librational effectons, increases. Big
macromolecules and small ions should have the opposite: positive and negative
effects on the stability and volume of primary librational effectons.

Macromolecules or polymers with molecular mass less than 2000 do not satisfy
the condition (7) and can not stimulate the growth of librational effectons.
On the other hand, a considerable increase in the concentrations of even big
polymers, when the average distance between them (eq.5) comes to be less than
the dimensions of a librational water effecton, perturbs clusterphilic
interactions and decreases freezing temperature, reducing water activity
(K\"{a}iv\"{a}r\"{a}inen, 1985, Fig.82).

\textbf{In general case each guest macromolecule has two opposite effects on
clusterphilic interactions. The equilibrium between these tendencies depends
on the temperature, viscosity, concentration of a guest macromolecule, its
dynamics and water activity.}

\textbf{When solute particles are sufficiently small they can associate due to
distant Van der Waals interaction, forming coherent primary effectons, when
[solute-solvent] interaction is more preferable than [solvent-solvent] and the
conditions (1 and 2) are fulfilled.}

\textbf{Important confirmation of this consequence of mesoscopic theory of
complex systems is the observation of compact clusters of ions even in dilute
salt solutions. For example, the extended x-ray absorption fine structure data
showed that the average distance between }$Zn^{2+}$\textbf{and }$Br^{-}%
$\textbf{\ is }$2.37\,\AA$\textbf{\ in 0.089M ZnBr}$_{2}$\textbf{\ aqueous
solutions and }$2.30\,\AA$\textbf{\ in 0.05 M solutions (Lagarde et al., 1980).}

These values are very close to the inter ionic distance observed in the
crystalline state $(2.40\AA)$.

The same conclusions was reached for NiBr$_{2}$ ethyl acetate solutions (Sadoc
et al., 1981) and aqueous CuBr$_{2}$ solutions (Fontaine et al.,1978). The
average distance between ions for the case of monotonic spatial distribution
(see eq.5) are much bigger than in ionic clusters under experimental conditions.

\textbf{Our theory of solutions consider formation of crystallites (inorganic
ionic clusters), as a solute coherent primary effectons self-organization. It
is more favorable process, than separate ion-water interactions.}

\textbf{For quantitative application of hierarchic theory to the description
of the processes in different solutions, using our special software, it is
necessary to have four experimental parameters, obtained under the same
conditions (temperature and pressure):}

\textbf{- density,}

\textbf{- sound velocity,}

\textbf{- positions of translational and librational bands in oscillatory spectra,}

\textbf{- refraction index.}

\textbf{The Combined analyzer of matter properties (CAMP), proposed by us (see
URL: http://kftt.karelia.ru/\symbol{126}alexk), could be the main tool for
such a measurements and study of complicated properties of solutions and
colloid systems.}

\bigskip

\begin{center}
{\Large The multi-fractional nature and properties of interfacial water, }

{\Large leading from hierarchic theory}
\end{center}

\smallskip

\textbf{We can present here a classification and description of four
interfacial water fraction properties, based on our hierarchic model:}

\textbf{1. The first fraction - primary hydration shell with maximum energy of
interaction with surface. The structure and dynamics of this 1st fraction can
differ strongly from those of bulk water. Its thickness: (3-10 \AA)
corresponds to 1-3 solvent molecule; }

\smallskip

\textbf{2. The second fraction - vicinal water (VW) is formed by elongated
primary lb effectons with saturated hydrogen bonds. It is a result of lb
effecton adsorption on the primary hydration shell from the bulk volume. The
thickness of this fraction of interfacial water: (30-75 \AA) corresponds to
10-25 molecules and is dependent on temperature, dimensions of colloid
particles and their surface mobility. VW can exist in rigid pores of
corresponding dimensions;}

\smallskip

\textbf{3. The third fraction of interfacial water - the surface-stimulated
Bose-condensate (SSBC), represented by 3D network of primary lb effectons has
a thickness of (50-300 \AA). It is a next hierarchical level of interfacial
water self-organization on the surface of second fraction (VW). The time of
gradual formation of this 3D net of linked to each other coherent clusters
(strings of polyeffectons), is much longer than that of VW and it is more
sensitive to temperature and other perturbations. The second and third
fractions of interfacial water can play an important role in biological cells activity;}

\smallskip

\textbf{4. The biggest and most fragile forth fraction of interfacial water is
a result of slow orchestration of bulk primary effectons in the volume of
primary (electromagnetic) lb deformons. The primary deformons appears as a
result of three standing IR photons (lb) interception. Corresponding IR
photons are superradiated by the enlarged lb effectons of vicinal water}.
\textbf{The linear dimension of primary IR deformons is about half of
librational IR photons, i.e. 5 microns (5}$\cdot10^{4}\,\AA)$\textbf{. This
''superradiation orchestrated water (SOW)'' fraction easily can be destroyed
not only by temperature increasing, ultrasound and Brownian movement, but even
by mechanical shaking. The time of spontaneous reassembly of this fraction
after destruction has an order of hours and is dependent strongly on
temperature, viscosity and dimensions of colloid particles. The processes of
self-organization of third (SSBC) and forth (SOW) fractions can be
interrelated by feedback interaction.}

\textbf{\smallskip}

\begin{center}
\textbf{\ Consequences and predictions of new model of interfacial solvent
structure}\smallskip

\smallskip
\end{center}

\ In accordance to generally accepted and experimentally proved models of
hydration of macromolecules and colloid particles, we assume that the first
2-3 layers of strongly bound water molecules, serves like intermediate shell,
neutralizing or ''buffering'' the specific chemical properties of surface
(charged, polar, nonpolar, etc.). Such strongly bound water can remain
partially untouched even after strong dehydration of samples in vacuum.

This first fraction of interfacial water serves like a matrix for second
fraction - vicinal water shell formation. The therm \textbf{'paradoxal
effect'}, introduced by Drost-Hansen (1985) means that the properties of
vicinal water are independent on specific chemical structure of the surface
from quartz plates, mineral grains and membranes to large macromolecules
(Clegg and Drost-Hansen, 1991). This can be a result of ''buffering'' effect
of primary hydration shell.

\smallskip

\textbf{Vicinal water (VW)} is defined as a water the structure of which is
modified in the volume of pores, by proximity to curved and plate interfaces
and interaction with strongly 'bound' water.

\textbf{For discussion of Vicinal water (VW) properties we proceed from the
statement that if the rotational correlation time of hydrated macromolecule is
less than that of primary librational effecton of bulk water, corresponding to
condition (1.3 and 1.4), these effectons should have a tendency to
''condensate'' on their hydration shell. It is a condition of their
''surviving'' and life-time increasing, because the resulting impulse of the
primary effectons is close to zero, in accordance to our model.}

\textit{The decreasing of most probable lb thermal impulses of water
molecules, especially in direction, normal to the surface of macromolecule or
colloid particle, should lead to increasing of corresponding edges of the 1st
''ground'' layer of primary lb effectons as compared to the bulk ones: }
\begin{equation}
\left[  \lambda=h/mv_{gr}\right]  _{lb}^{vic}>\,\left[  \lambda=h/mv_{gr}%
\right]  _{lb}^{bulk}\tag{20b}%
\end{equation}

This turns the cube-like shape of effectons of the bulk water to shape of
elongated parallelepiped..

Consequently, the volume of these waves B three-dimensional (3D)
superposition, representing \textit{the vicinal} \textit{lb }and \textit{tr
primary }effectons - is bigger than that of bulk liquid effectons. As far the
stability and life-time of these enlarged primary lb effectons are increased,
it means the increasing of their concentration as well.

As far we assume in our multifractional model, that \textbf{VW} is a result of
''condensation'' of primary librational effectons on primary hydration shell
and their elongation in direction, normal to surface, we can make some
predictions, related to properties of this 2nd interfacial water fraction:

1. The thickness of VW can be about 30-75 \AA, depending on properties of
surface (geometry, polarity), temperature, pressure and presence the
perturbing solvent structure agents;

2. This water should differ by number of physical parameters from bulk water.
It can be characterized by:

a) lower density;

b) bigger heat capacity;

c) bigger sound velocity

d) bigger viscosity;

e) smaller dielectric relaxation frequency, etc.).

\smallskip

\textbf{These differences should be enhanced in a course of third fraction of
interfacial water formation: surface-stimulated Bose-condensate (SSBC) as far
the concentration of primary librational effectons in this fraction is bigger
than that in bulk water.} The time, necessary for SSBC three-dimensional
structure self-organization can have an order of minutes or hours, depending
on temperature, geometry of surface and average distance between interacting surfaces.

\smallskip

From Fig.4b we can see that the linear dimension of primary librational
effecton of bulk water at 25$^{0}C\,$ is about $\left[  \lambda\right]
_{lb}^{bu}\symbol{126}$ 15\AA. The lower mobility of water molecules of
vicinal water is confirmed directly by almost 10-times difference of
dielectric relaxation frequency (2$\cdot10^{9}$ Hz) as respect to bulk one
(19$\cdot10^{9}\,Hz)\;$(see Clegg and Drost-Hansen, 1991). The consequence of
less mobility and most probable impulse (momentum) of water molecules should
be the increasing of most probable wave B length and dimensions of primary
effectons. \textbf{The enhancement of lb primary effecton edge should be more
pronounced in the direction, normal to the interface surface. In turn, such
elongation of coherent cluster can be resulted in increasing the intensity of
librational IR photons superradiation.}

\textbf{\smallskip}

The increasing of temperature should lead to decreasing the vicinal
librational effectons dimensions.

The dimensions of primary translational effectons of water is much less than
of librational ones (see Fig. 4a). The contribution of translational effectons
in vicinal effects is correspondingly much smaller than of librational.

Our model predicts that not only dimensions but as well concentrations of
primary librational effectons should increase near rigid surfaces. The
\textit{system of vicinal 3D polyclustrons }can emerge.

\textbf{In accordance to our model the Drost-Hansen thermal anomalies of
vicinal water behavior near }$15^{0},\,30^{0},45^{0}\,$\textbf{and 60}$^{0}%
\;$\textbf{has the same explanation as presented in comments to Fig. 4a
\thinspace for bulk water. Because the dimensions, stability and concentration
of vicinal librational effectons are bigger than that of bulk effectons the
temperature of anomalies for these two water fraction can differ also.}

\textbf{As far the positions of disjoining pressure sharp maxima of water
between quartz plates did not shift markedly when their separation change from
100 \AA\ to 500 \AA\ it points that\ the main contribution in temperature
anomalies is related to bulk librational effectons. The same is true for
viscosity measurements of water between plates with separation: 300-900 \AA.}

\textbf{The possible explanation of Drost-Hansen temperature values stability
is\ that the vicinal water layer (50-70\AA) has the bigger dimensions of
primary librational effectons edge, than that of bulk water, presumably only
in the direction, normal to the surface of the interface. Such first layer of
surface - stimulated Bose-condensation induces the formation of string-like
librational polyeffectons, composed from elongated in the same direction bulk
primary effectons and stabilized by Josephson's junctions and distant
Van-der-Waals interactions. These strings can be cross-linked by short
''chains'' of primary lb effectons, making such 3D polyeffecton net more
stable as respect to Brownian motion. We can term this fraction of interfacial
water as ''Surface-stimulated Bose-condensate (SSBC)''. As far in such
mesophase the linear dimensions of two of three edges of primary lb effectons,
approximated by parallelepiped, remains the same as in the bulk water, it
explains the relative stability of thermal anomalies at temperature scale. It
means the similar conditions of maximum stability for primary lb effectons
(see eq.20c), in composition of SSBC and calculated for bulk water. }

For the other hand, the elongated structure of primary lb effectons, composing
SSBC, should increase all the effects, related to intensive and coherent
superradiation of lb IR photons in directions of the longest edge of the
effectons (see Introduction). \textbf{This largest and most subtle fraction of
interfacial water also has more orchestrated cooperative properties than the
unperturbed by superradiation bulk water due to stronger distant Van-der-Waals
interaction between primary lb effectons. Such a ''superradiation-orchestrated
water (SOW)'' is even less ordered and stable than the second fraction of
interfacial water: surface-stimulated Bose-condensate (SSBC).}

\textbf{\ Its 3D structure of SSBC and SOW could be easily destroyed by
mechanical perturbation or heating. The relaxation time of ''regeneration'' or
self-organization of these water fractions can take an hours and depends on
temperature and mobility of the surface. }

\smallskip

\textbf{The sharp conditions of maximum stability }\textit{\ of librational
primary effectons at certain \textbf{temperatures - the integer number of
molecules in the edge of lb effecton: }}
\begin{equation}
\kappa=\text{\textit{\textbf{6, 5, 4, 3, 2\ \ \ \ (see comments to Fig.4a)}}%
}\tag{20c}%
\end{equation}
\textit{\textbf{\ } are responsible for deviations of temperature dependencies
for many parameters of water from monotonic ones.}

\smallskip

\textbf{\ Vicinal water (VW) and surface-stimulated Bose-condensation of water
near biological membranes and filaments (microtubules, actin polymers) can
play an important regulation role in cells and their compartments dynamics and
function. Its highly cooperative properties and thermal sensitivity near
Drost-Hansen temperatures can be used effectively in complicated processes,
related to cells proliferation, differentiation and migration. Changing of
water activity: its increasing as a result of VW and SSBC deassembly and
decreasing in a course of it 3D structure self-organization, can affect
strongly the dynamic equilibrium [association }$\rightleftharpoons
\,$\textbf{dissociation}$]$\textbf{\ of oligomeric proteins, their allosteric
properties and osmotic processes in cells.}

\ 

\medskip

\begin{center}
\textbf{COMPARISON OF EXPERIMENTAL DATA WITH THEORETICAL PREDICTIONS}

\medskip
\end{center}

\ 

\textbf{It will be shown below that our hierarchic model of interfacial water
explains the comprehensive and convincing experimental data, available on subject.}

\textbf{The following experimental results, illustrating the difference
between the second (VW) and third (SSBC) fractions of interfacial and bulk
water will be discussed:}

\textbf{1. The lower density of vicinal water near plates and in pores
(\symbol{126}0.970g/cm}$^{3})\,\,$\textbf{as compared to bulk one
(\symbol{126}1.000 g/cm) (Etzler and Fagundus, 1983; Low, 1979; Clegg 1985);}

\textbf{2. Different selectivity of vicinal water in pores to
structure-breaking and structure-making inorganic ions; }

\textbf{3. Volume contraction on sedimentation of particles dispersed in water
(Braun et al. 1987);}

\textbf{4. Higher heat capacity of vicinal water as compared to bulk water
(Etzler 1988);}

\textbf{5. Higher ultrasound velocity in the interfacial water;}

\textbf{6. Higher ultrasound absorption in the interfacial water;}

\textbf{7. Higher viscosity of interfacial water (Peshel and Adlfinger 1971)
and its dependence on shearing rate;}

\textbf{8. Sharp decreasing of the effective radius of dilute solution of
polysterone spheres, in a course of temperature increasing.}

\ 

\smallskip

In accordance to our model, the water molecules in composition of primary
librational effectons (see Fig. 4a) are four-coordinated like in ideal ice
structure with lowest density. In contrast to that, the water in the volumes
of translational effectons, [lb/tr] convertons and \textit{lb} and \textit{tr}
macroeffectons and supereffectons has the nonsuturated hydrogen bonds and a
higher density. The compressibility of primary [lb] effectons should be lower
and sound velocity - higher than that of bulk water. It is confirmed by
results of Teixeira et. al. (1985), obtained by coherent- inelastic- neutron
scattering. They point on existance in heavy water at 25 $^{0}$C the
solid-like collective excitations with bigger sound velocity than in bulk
water. These experimental data can be considered as a direct confirmation of
our primary librational effectons existence..

\smallskip

As far the fraction of water involved in primary librational effectons in
vicinal water is much higher than in the bulk one, this explains the result
$\left[  1\right]  $ at the list above. The biggest decreasing of density
occur in pores, containing enlarged primary librational effectons, due to
stronger water molecules immobilization.

\smallskip

Different selectivity of vicinal water in pores of silica gel (result [2]) to
structure-breaking and structure-making inorganic ions (Wiggins 1971, 1973),
leading to higher concentration of the former (like K$^{+})$ as respect to
latter (like Na$^{+})\;$ones in pores was revealed. It is in total accordance
with our model as far for penetration into the pore the ion have at first to
break the ordered structure of enlarged librational effectons in the volume of
pore. Such kind of Na/K selectivity can be of great importance in the passive
transport of ions throw the pores of biological membranes.\smallskip

Result $\left[  3\right]  \,\,$of volume contraction of suspension of 5-$\mu
m\,$silica particles in a course of their sedimentation - is a consequence of
mechanical perturbation of cooperative and unstable 3D fraction: surface -
stimulated Bose condensation (SSBC), its partial 'melting' and increasing of
water fraction with nonsuturated hydrogen bonds and higher density.

\smallskip

The available experimental data of the vicinal water thickness evaluate it as
about 50-70 \AA\ (Drost-Hansen, 1985). In totally or partly closed volumes
like in silica pores the vicinal effect must be bigger than near the plain
surface. This explains the maximum heat capacity of water at 25$^{0}\,$(result
$\left[  4\right]  )$ in silica pores with radii near 70 \AA\ (Etzler and
White, 1988).

As far the cooperative properties of 2nd and 3d fractions of interfacial
water, corresponding to vicinal water and surface - stimulated Bose
condensation are higher than that of bulk water, it explains the bigger heat
capacity of both of these fractions.

\smallskip

The higher sound velocity in the VW and SSBC fractions as compared to bulk
water (result [5]) is a consequence of higher concentration of primary
librational effectons with low compressibility due to saturated H-bonds.

\smallskip

Higher absorption of ultrasound by interfacial water (result [6]) can be a
consequence of dissipation processes, accompanied the destruction of water
fraction, corresponding to SSBC by ultrasound.

\smallskip

The higher viscosity of vicinal water (result $\left[  7\right]  )$ is a
result of higher activation energy of librational macrodeformons excitation
$\left(  \varepsilon_{D}^{M}\right)  _{lb}$ in a more stable system of vicinal
polyclustrons (see our mesoscopic theory of viscosity).

\smallskip

\textbf{Sharp decreasing of the effective dimensions of dilute solution of
polysterone (PS) spheres (0.1\%), in a course of temperature increasing
(result [8]) - is a consequence of cooperative destruction (melting) of 3d
fraction of interfacial water (SSBC).}

The corresponding transition occur at 30-34 $^{0}$C, as registered by Photon
correlation spectroscopy and is accompanied by the effective Stocks radius of
PS spheres decreasing onto 300 \AA.

\textbf{The ''regeneration time'' of this process is about 20 hours. It may
include both: time of SAPS self-organization and self-organization of the less
stable forth interfacial fraction: ''Water, orchestrated in the volume of IR
primary deformons''. }

\smallskip

The frequent non-reproducibility of results, related to properties of
interfacial water, including Drost-Hansen temperature anomalies, can be
resulted from different methods of samples preparation and experimental conditions.

For example, if samples where boiling or strongly heating just before the
experiment, the 3d and 4th fractions of interfacial water can not be
observable. The same negative result is anticipated if the colloid system in
the process of measurement is under stirring or intensive ultrasound radiation influence.

\begin{center}
\medskip

{\large Nonobserved yet predictions, related with third (SSBC) and forth
(water, orchestrated by IR standing photons) fractions of new interfacial model}
\end{center}

{\large \medskip}

1. Gradual increasing of pH of distilled water in a course of these fractions
formation near nonpolar surface - due to enhancement of probability of
superdeformons excitation. Corresponding increasing of concentration of
cavitational fluctuations are accompanied by dissociation of water molecules:
\[
H_{2}O\rightleftharpoons H^{+}+HO^{-}
\]

and the protons concentration elevation;

\smallskip

2. Increasing the UV and visible photons spontaneous emission near nonpolar
surface as a result of increasing of the frequency of water molecules
recombination:
\[
H^{+}+HO^{-}\rightarrow H_{2}O+photons
\]

These experiments should be performed in dark box, using sensitive photon
counter or photo-film.

\smallskip

The both predicted effects should be enhanced in system, containing parallel
nonpolar multi-layers, with distance ($l)$ between them, corresponding to
conditions of librational IR photons standing waves formation:
$l=5,\,10,\,15,\,20\,\,$microns.

\smallskip

3. More fast and ordered spatial self-organization of macromolecules, like
described in the next section is anticipated also in the volume of forth
fraction of interfacial water.

The dynamics of such process can be registered by optical confocal or
tunneling microscope.

\smallskip

4. Our model predicts also that the external weak coherent electromagnetic
field, generated by IR laser, like the internal one, also can stimulate a
process of self-organization in colloid systems.

\medskip

\begin{center}
\smallskip

{\large 4. Distant solvent-mediated interaction between macromolecules}
\end{center}

\smallskip

The most of macromolecules like proteins can exist in dynamic equilibrium
between two conformers (A and B) with different hydration $(n_{H_{2}O})$ and flexibility:%

\begin{equation}
A+n_{H_{2}O}\Leftrightarrow B\tag{11 a}%
\end{equation}

Usually correlation time of more hydrated B- conformer ($\tau_{B})$ and
its\textbf{\ effective} volume are less than that of A-conformer ($\tau_{A}):$%

\[%
\begin{array}
[c]{c}%
\tau_{A,B}=\frac{V_{A,B}}k(\eta/T)\\
\\
\tau_{A}>\tau_{B}\text{ \ \ and\ \ \ }V_{A}>V_{B}%
\end{array}
\]
This means that flexibility, determined by large-scale dynamics of B-conformer
is more than that of A-conformer.

For such case a changing of the bulk water activity $(a_{H_{2}O})$ in solution
by addition of another type of macromolecules or by inorganic ions induce the
change of the equilibrium constant: $K_{A\Leftrightarrow B}%
=(K_{B\Leftrightarrow A})^{-1}$ and the dynamic behavior of macromolecules
(K\"aiv\"ar\"ainen, 1985, 1986) :%

\begin{equation}
\Delta\ln K_{B\Leftrightarrow A}=n_{H_{2}O}\cdot\Delta\ln a_{H_{2}O}\tag{12}%
\end{equation}

In mixed systems: [PEG + spin-labeled antibody] the dependence of
\textbf{large-scale (LS) }dynamics of antibody on the molecular mass of
polyethylenglycol (PEG) is similar to dependence of water activity and
freezing temperature $(T_{f})$ of PEG solution, discussed above
(K\"aiv\"ar\"ainen, 1985, Fig. 82).

The presence of PEG with mass and concentration increasing$\;(a_{H_{2}O})$ and
$(T_{f})$ stimulate the LS-dynamics of proteins decreasing their effective
volume $V$ and correlation time $\tau_{M}$ in accordance to eq.(12).

If $\;\Delta T_{f}=T_{f}^{0}-T_{f}\;$ is the difference between the freezing
point of a solute $(T_{f}^{0})$ and solution $(T_{f})$, then the relation
between \textbf{water activity in solution} and $\Delta T_{f}$ is given by the
known relation:%

\begin{equation}
\ln a_{H_{2}O}=-\left[
\begin{array}
[c]{c}%
\Delta H/R\left(
\begin{array}
[c]{c}%
T_{f}^{0}%
\end{array}
\right)  ^{2}%
\end{array}
\right]  \cdot\Delta T_{f}\tag{13}%
\end{equation}

where $\Delta$H is the enthalpy of solute (water) melting; R is the gas constant.

In our experiments with polymer solutions 0.1 M phosphate buffer $(pH7.3+0.3M
$ NaCl) was used as a solvent (K\"aiv\"ar\"ainen, 1985, Fig. 82).

One can see from (13) that the negative values of $\Delta T_{f}$ in the
presence of certain polymers means increasing water activity in three
component [water - ions - polymer] system $(\Delta\ln a_{H_{2}O}>0)$. In turn,
it follows from (12) that the $\left[  A\Leftrightarrow B\right]  $
equilibrium of guest proteins in the same system will shift to the right.
Consequently, the flexibility of the proteins will increase. Correlation
between $T_{f}$, water activity $(a_{H_{2}O})$ and immunoglobulin flexibility
$(\tau_{M})$, corresponding to (11 - 13) was confirmed (K\"{a}iv\"{a}%
r\"{a}inen, 1985, Table 13).

\textbf{It was shown in our experiments that protein-protein distant
interaction depends on their large-scale (LS) dynamics and conformational
changes induced by ligand binding or temperature (K\"{a}iv\"{a}r\"{a}inen,
1985). Such distant solvent-mediated effects may be explained using our idea
of inter- and intramolecular clusterphilic interactions, discussed above.}

We use the assumption that between inter- and intra-clusterphilic interaction
the positive correlation exist. It means that increase of dimensions and
\textit{stabilization of the librational bulk (inter) water effectons induce
an increase of the water clusters dimensions in protein cavities }and shift
$A\Leftrightarrow B$ equilibrium of the cavities to the right, i.e. to the
more flexible conformer. This leads to decreasing of resulting correlation
time and effective volume of protein.

Our interpretation is confirmed by fact that a decline in temperature,
increasing the dimensions of the bulk librational effectons has the
stimulating influence on the flexibility of protein like the presence of guest
macromolecules. \textit{Decreasing of the temperature shifts the equilibrium
between hydrophobic and clusterphilic interaction to the latter one. }

At \textit{low concentration of macromolecules }$(C_{M})$, when the average
distance (r) between them (eq.5) is much more than dimensions of primary
librational water effecton $(r\gg\lambda_{lb})$, the intermolecular
clusterphilic interaction does not work effectively. In this case the
large-scale $\left[  A\Leftrightarrow B\right]  $ pulsations of proteins,
accompanied by \textbf{acoustic impulses in solvent} can enhance the
\textit{activity of water}.

This dynamic effect of proteins on solvent can be responsible for the distant
solvent- mediated interaction between macromolecules at low concentration
(K\"aiv\"ar\"ainen, 1985, 1987).

Acoustic impulses in protein solutions are result of the jump-way
$B\rightarrow A$ transition of interdomain or intersubunit cavities with
characteristic time about $10^{-10}\sec$. This rapid transition follows the
cavitational fluctuation of a water cluster formed by 30 - 50 water molecules.
The fluctuation is a result of conversion of (\textit{lb}) primary effecton to
(\textit{tr}) one: $[lb/tr]$ converton excitation.

\textbf{The hydrodynamic Bjorkness interaction }between different type and
identical proteins can be induced by acoustic wave packets in solvent,
stimulated by large-scale pulsations of proteins (K\"aiv\"ar\"ainen et al.,
1988). This new approach was used for estimation of frequency of LS-
pulsations of interacting proteins like (immunoglobulins) as $\left(
10^{4}-10^{5}\right)  $s$^{-1}$.

In very concentrated solutions of macromolecules, when the distance between
macromolecules starts to be less than linear dimension of primary librational
effecton of water:$\;$

$r$ $\le\lambda_{lb},$ the trivial aggregation process begins to dominate. It
is related to the decrease of water activity in solutions.

Let us analyze in more detail the new effect - increasing water activity
$(a_{H_{2}O})$ under the effect of macromolecules in a three component [water
- salt - macromolecules] system. The Gibbs-Duhem law for this case can be
presented as (K\"{a}iv\"{a}r\"{a}inen, 1988):%

\begin{equation}
X_{H_{2}O}\Delta\ln a_{H_{2}O}+X_{M}{\frac{\Delta\bar{\mu}_{M}}{RT}}%
+X_{i}\Delta\ln a_{i}=0\tag{14}%
\end{equation}

where $X_{H_{2}O},\;X_{M},\;X_{i}$ are the molar fractions of water,
macromolecules and ions in the system;%

\begin{equation}
a_{j}=y_{j}X_{j}=\exp\left(  -{\frac{\mu_{0}-\mu_{j}}{RT}}\right)
=\exp\left(  -{\frac{\Delta\mu_{j}}{RT}}\right) \tag{15}%
\end{equation}
is the activity of each component related to its molar fraction (X$_{j}$) and
coefficients of activity (y$_{j}$);%

\begin{equation}
\Delta\mu_{M}\simeq(G_{B}-G_{A})\Delta f_{B}\tag{16}%
\end{equation}
- the change of the mean chemical potential $(\bar{\mu}_{M})$ of a
macromolecule (protein) pulsing between A and B conformers with corresponding
partial free energies $\bar{G}_{A}$ and $\bar{G}_{B}$, when the change of
B-fraction is $\Delta f_{B}$ and%

\begin{equation}
\mu_{M}\cong f_{B}G_{B}+f_{A}G_{A}\tag{17}%
\end{equation}%
\begin{equation}
\Delta\ln a_{i}=(\Delta a_{i}/a_{i})\simeq-\Delta\kappa_{i}\tag{18}%
\end{equation}
where%

\begin{equation}
\kappa_{i}=1-y_{i}\tag{19}%
\end{equation}
is the fraction of thermodynamically excluded ions (for example, due to ionic
pair formation).

One can see from (15) that when $a_{H_{2}O}<1$, it means that%

\begin{equation}
\mu_{H_{2}O}^{0}>\mu_{H_{2}O}^{S}=H_{H_{2}O}^{S}-TS_{H_{2}O}^{S}\tag{20}%
\end{equation}
It follows from (20) that the decrease of water entropy (\={S}) in solution
related to hydrophobic and clusterphilic interactions may lead to increased
$\mu_{H_{2}O}^{S}$ and water activity.

It is easy to see from (14) that the elevation of concentration and $X_{M}$ of
macromolecules in a system at constant temperature and $\Delta\bar{\mu}_{M}$
may induce a rise in water activity $(a_{H_{2}O})$ only if the activity of
ions $(a_{i}=y_{i}X_{i})$ is decreased. The latter could happen due to
increasing of fraction of thermodynamically excluded ions $(\kappa)\;($eqs. 18
and 19).

There are \textit{two processes }which may lead to increasing the probability
of ionic pair formation and fraction $\kappa$ elevation.

The \textit{first }one is the forcing out of the ions from the ice-like
structure of enlarged librational effectons, stimulated by the presence of
macromolecules. This exclusion volume phenomenon increases the effective
concentration of inorganic ions and their association probability.

The \textit{second }one dominates at the low concentration of $\left[
A\Leftrightarrow B\right]  $ pulsing macromolecules, when thixotropic
structure fail to form $(r=11.8/C_{M}^{1/3}\gg\lambda_{lb})$. Acoustic
impulses in solvent generated by pulsing proteins stimulate the fluctuation of
ion concentration (K\"{a}iv\"{a}r\"{a}inen, 1988) increasing the probability
of ionic pairs formation.

\begin{center}
{\large 5. Spatial self-organization in the water-macromolecular systems}
\end{center}

\smallskip

The new type of self-organization in aqueous solutions of biopolymers was
revealed in Italy (Giordano et al., 1981). The results obtained from
viscosity, acoustic and light-scattering measurements showed the existence of
long-range structures that exhibit a \textit{thixotropic behavior}. This was
shown for solutions of lysozyme, bovine serum albumin (BSA), hemoglobin and
DNA. Ordered structure builds up gradually in the course of time to become
fully developed after more than 10-15 hours.

When a sample is mechanically shaken this type of self-organization is
destroyed. The ''preferred distance'' between macromolecules in such an
ordered system is about $L\simeq50\,\AA$ as revealed by small angle neutron
scattering (Giordano et al., 1991). \textbf{It is important that this distance
can be much less than the average statistical distance between proteins at low
molar concentrations} $\mathbf{(}\mathbf{C}_{\mathbf{M}}\mathbf{)\;}($see eq.5).

\textbf{This fact points to attraction force between macromolecules. In
accordance with our mesoscopic model, this attraction is a result of external
intermolecular clusterphilic interaction. }It has been shown experimentally
that hypersonic velocity in the ordered thixotropic structures of 10\%
lysozyme solutions is about $2500m/s$, i.e. 60\% higher than that in pure
water $(1500m/s)\,\,$(Aliotta et. al., 1990) Quasielastic and elastic neutron
scattering in 10\% lysozyme aqueous system at $20^{0}$ shows that the dynamic
properties of the ''bound'' water in a thixotropic system are similar to the
properties of pure water at $3-4^{0}C$, i.e. $16^{0}C$ below actual
temperature (Giordano et al., 1991). Experimental evidence for heat capacity
increasing in lysozyme solutions during 10-15 hours of self-organization was
obtained by adiabatic microcalorimetry (Bertolini et al., 1992). The character
of this process is practically independent on pH and disappears only at the
very low concentration of protein $(<0.2\%)$, when the average distance
between macromolecules becomes too big. Increasing the temperature above
$40^{0}$ also inhibits thixotropic-type of self-organization in
water-macromolecular systems.

In the series of experiments with artificial polymers - polyethyleneglycols
(PEG) with decreasing molecular mass it was shown that self-organization (SO)
in 0.1 M phosphate buffer $(pH\,7.3,\;25^{0})$ exists in the presence of PEG
with a weight of 20.000, 10.000 and 8.000 daltons, but disappears at a
molecular mass of 2.000 and lower (Salvetti et al., 1992).

These data are in good agreement with the influence of PEG on the freezing
temperature and intramolecular interaction described above.

Independently of Italian group, similar ordering process were observed in
Japan (Ise and Okubo, 1980) for aqueous solutions of macroions.

The process of ''compactization'' or ''clusterization'' of solute (guest)
molecules in one volumes of solution or colloid system, should lead to
emergency of voids in another ones. Such a phenomena were revealed by means of
confocal laser scanning microscope, ultramicroscope and video image analyzer
(Ito et al.,1994; Yoshida et al.,1991; Ise et al.,1990).

Inhomogeneity of guest particles distribution were revealed in different ionic
systems, containing ionic polymers or macroions, like sodium polyacrilate, the
colloid particles, like polysterone latex $(N\,300,1.3\,\mu C/cm^{2})$ and
Langmuir-Blodgett films. The time evolution of the numbers of different
clusters from such particles were followed during few hours.

\smallskip

\textbf{The colloid crystal growth at 25 }$^{0}$\textbf{C in }$H_{2}%
O$\textbf{\ and }$(H_{2}O-D_{2}O)$\textbf{\ systems can be divided on four
stages (Yoshida et all.,1991).}

\textbf{In the first stage the particles were diffusing freely.}

\textbf{In the second stage the local concentration of the particles took place.}

\textbf{In the third stage clusters from 3-10 particles were formed.}

\textbf{In the last fourth stage the smaller clusters turns to the bigger ones
and macroscopically well ordered structures were formed. Simultaneously the
huge voids as large as }$50-150\mu m$\textbf{\ were observed.}

Such an ordering can be immediately destroyed by mechanical shaking or by
adding of inorganic salts (NaCl), even in such relatively small concentrations
as $10^{-4}$ mol$/dm^{3}$.

Authors conclude that the repulsion as only one assumption is not enough for
explanation of the phenomena observed. The model, considering a
\textit{short-range repulsive }interaction and \textit{long-range attraction
}between particles should be used. They try to explain attraction between
similar charged macroions by presence of small inorganic counterions. Authors
believe that short range repulsion can be overwhelms by attraction. But such a
simple model does not explain ''clusterization'' of electrically neutral guest
macromolecules and acceleration of this process due to decreasing the
concentration of inorganic salts in presence of ionic sorbents.

Despite a large amount of different experimental data and great importance,
the mechanism of spontaneous type of self-organization in colloid systems
remains unknown. It is evident, however, that it can not be attributed to
trivial aggregation.

Our explanation of distant attraction between macromolecules, by means of
clusterphilic interaction and polymerization of clustrons in the volume of
primary electromagnetic deformons (see below) is more adequate for description
of self-organization phenomena in aqueous colloid systems, than counterion hypothesis.

\textbf{Just clusterphilic interactions determines the attraction between
large guest molecules or colloid particles on mesoscopic scale. They are
responsible for the lot of vicinal water effects described below.}

Our computer calculations of water properties based on the mesoscopic theory
show that the volume of primary librational effecton of water at $25^{0}$
includes about 100 water molecules (Fig. 7a of [1] or Fig. 4a of [2]).

In accordance with condition (7) for clusterphilic interactions, the
macromolecules with volume
\[
V_{M}>100\cdot(V_{0}/N_{0})
\]
can decrease the librational mobility of $H_{2}O$ molecules, their impulses
and, consequently, (eqs. 1 and 9) increase the dimensions of the librational
effectons of water. Obviously, there must exist a direct correlation between
the effective Stokes radius of a macromolecule (i.e. its molecular mass) and
the process of self-organization.

\textbf{\ The mass of a lysozyme (Lys) is about 13.000 D, and available
experimental data (Giordano et al., 1991) show that\ the mobility of water in
a hydration shell in composition of thixotropic system at }$25^{0}%
$\textbf{\ is about 3 times less than that of pure water. This mobility is
directly related to the most probable group velocity and impulse of }%
$H_{2}O\,\,(mv_{gr})$\textbf{. It means a three-fold increase in the
dimensions of librational effectons in the presence of lysozyme (see eq. 1).}%

\begin{equation}
\lbrack\lambda_{H_{2}O}\approx15\,\AA]\rightarrow[L_{(H_{2}O}\approx
45\,\AA]\tag{20a}%
\end{equation}
This value is quite close to the experimental \textit{preferred distance
}$(50\overset{o}{A})$ between proteins after self-organization (SO). Because
the shape of the effectons can be approximated by parallelepiped or cube, we
suggest that each of its 6 sides can be bordered and stabilized by at least
one macromolecule (Table). We termed \textit{corresponding type of
quasiparticles ''clustrons''.} Enlarged librational effectons serve as a
''glue'', promoting interaction between surrounding macromolecules or colloid
particles. Probability of librational effectons dissociation to translational
ones, i.e. $[lb/tr]$ convertons excitation,- is decreased in composition of clustrons.

Cooperative water clusters in the volume of clustrons are very sensitive to
perturbation by temperature or by ion-dipole interactions. When the colloid
particles have their own charged groups on the surface, very small addition of
inorganic salts can influence on clustrons formation.

\smallskip

{\large Our model can answer the following questions:}

\textbf{1. Why polymers with a molecular mass less than 2000 D do not
stimulate self-organization ?}

\textbf{2. Why is self-organization inhibited at sufficiently high temperature
}$(>40^{0})$\textbf{\ ?}

\textbf{3. What causes the repulsive hydration force?}

\smallskip

The answer to the \textbf{first question }is clear from condition (7) and
eqs.(8). Small polymers can not stimulate the growth of librational effectons
and clustrons formation due to their own high mobility.

The response to the \textbf{second question }is that a decrease in the
dimensions of librational effectons with temperature (Fig. 7 of [1]) leads to
deterioration of their cooperative properties and stability. The probability
of clustrons formation drops and thixotropic structures can not develop.

The \textbf{repulsion, }like attraction between colloid particles can be
explained by clusterphilic ineraction.. The increasing of external pressure
should shift the $(a\Leftrightarrow b)_{lb}$ equilibrium of librational water
effectons to the left, like temperature decreasing. This means the enhancement
of librational water effecton (and clustron) stability and dimensions. For
this reason a \textit{repulsive or disjoining hydration force} arises in
different colloid systems. The swelling of clays against imposed pressure also
is a consequence of clusterphilic interactions enhancement, i.e. enlargement
of water librational effectons due to $H_{2}O$ immobilization.

In accordance with our model, effectons and their derivatives,
\textit{clustrons}, are responsible for self-organization in colloid systems
on a \textit{mesoscopic }level (Stage II in Table).

The \textit{macroscopic }level of self-organization (polyclustrons),
responsible for the increase of viscosity and long-distance interaction
between clustrons originates due to interaction between the electric and
magnetic dipoles of clustrons (Stage III, Table). Because the thermal movement
of the $H_{2}O$ molecules, composing clustrons is coherent, the \textit{dipole
moment of clustron} is proportional to the number of $H_{2}O$ in its volume.

\begin{center}%
\begin{center}
\includegraphics[
height=5.0825in,
width=4.3215in
]%
{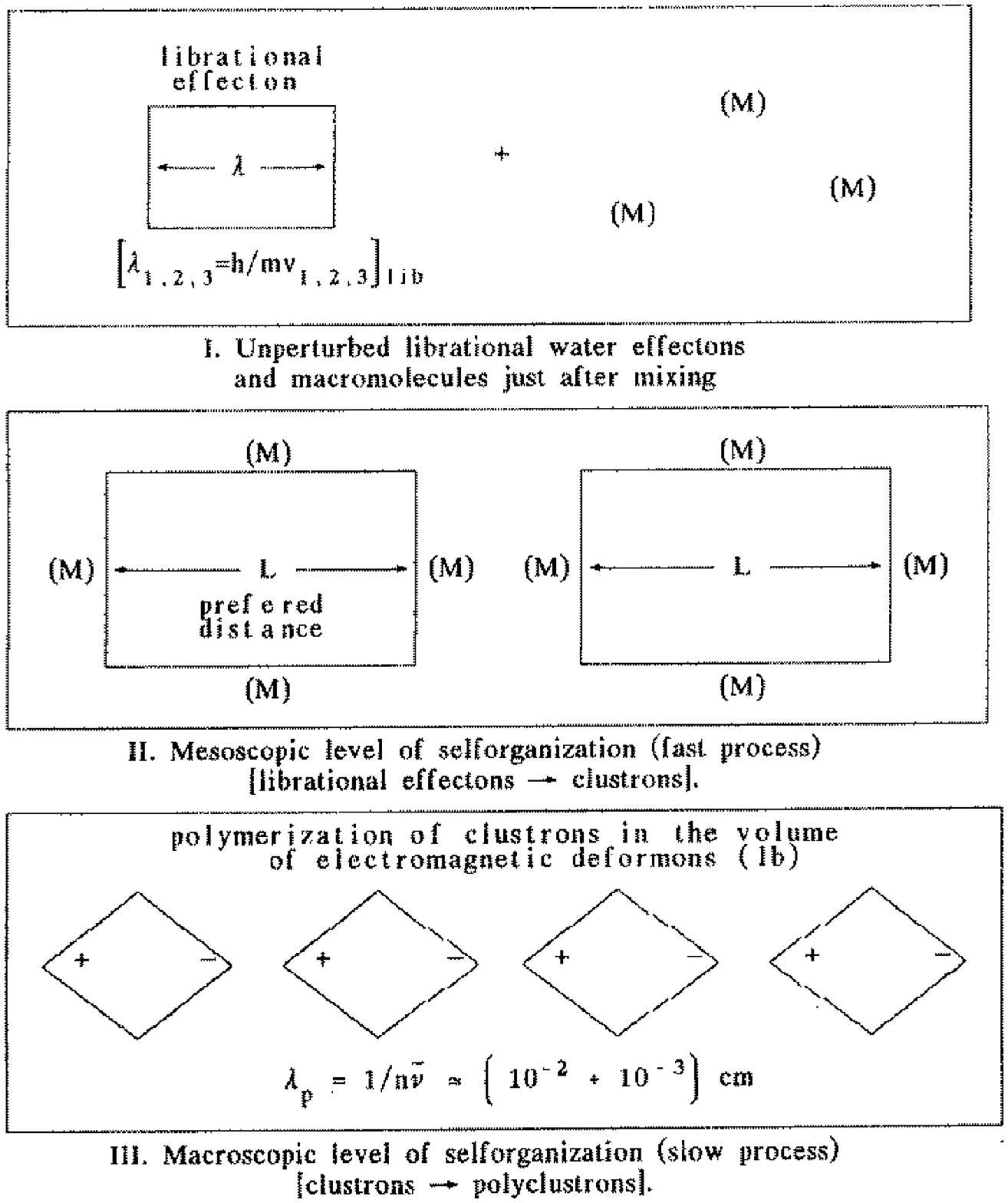}%
\end{center}
\end{center}

\begin{quotation}
\textbf{Table. }Schematic representation of three stages $(I\rightarrow
II\rightarrow III)$ of gradual spatial self-organization in aqueous solutions
of macromolecules $\left(  M\right)  $.
\end{quotation}

\smallskip

\textbf{\ Formation of polyclustrons (Stage III, Table) as the space-time
correlated large group of clustrons can be stimulated in the volume of primary
electromagnetic deformons (tr and lib), having almost macroscopic dimensions.}

\textbf{Clustrons (stage II) are complexes of enlarged primary librational
effectons, bordered by each of 6 side with macromolecules. The increase of
librational effecton dimensions }$(\lambda\rightarrow L)$\textbf{\ is related
to decreased water molecule mobility due to external clusterphilic interaction.}

\textbf{Polyclustrons (stage III) are space and time correlated systems of a
large group of clustrons in the volume of electromagnetic deformons. Their
linear dimensions }$(0.01-0.001cm)$\textbf{\ correspond to translational and
librational wave numbers:}%

\[
\tilde{\nu}_{tr}\simeq60cm^{-1}\text{ \ \ and\ \ \ }\tilde{\nu}_{lb}%
\simeq700cm^{-1}
\]
We suppose that for optimal process of self-organization, the molar
concentration of macromolecules $(C_{M})$ must satisfy the condition:%

\[
10L>[r=11.8\cdot C_{M}^{-1/3}]\ge L,
\]

where: $L\cong q\lambda_{lb}=h/m(v_{lb}/q)$ is the dimension of a clustron
$(\lambda_{lb}^{1,2,3}=h/mv_{lb}^{1,2,3}$ is the dimension of primary
librational effecton in pure water; $r$ - statistically most probable distance
between macromolecules $(eq.5);\;q$ is the \textit{lib} effecton magnification
number, reflecting the effect of water molecules immobilization (decreasing of
their group velocity) in presence of macromolecules or colloid particles.

Under condition $r>>10L$, formation of \textit{clustrons }can be accompanied
by big voids in colloid system. At the opposite limit condition $r<L$, the
trivial aggregation will dominate, accompanied by ''melting'' of vicinal
librational effectons.

The librational IR photon wave length, calculated from the oscillatory spectra
of water, is about (eq. 1.37):%

\[
\lambda_{p}=(n\cdot\tilde\nu)^{-1}\approx(1.33\cdot700)^{-1}\approx0.001cm
\]
where $\tilde\nu\simeq700\,cm^{-1}$ is the wave number of librational band in
the oscillatory spectra of water; $\;n=1.33$ is the water refraction index.

The 3D superposition (interception) of such three photons forms primary
electromagnetic deformons, stimulating development of polyclustron system.

\textbf{Our model predicts that the external electromagnetic field, like
internal one, also can stimulate a process of self-organization in colloid
systems. On the other hand, the agents perturbing the structure of librational
effectons (temperature, ions, organic solvents) should have the opposite
effect. The slowest stage of self-organization, polyclustrons and polyclustron
net formation (thixotropic -like structure), is also sensitive to mechanical
shaking. Due to the collective effect the energy of interaction between
clustrons in polyclustrons is higher than kT at sufficiently low temperatures.}

\smallskip

\textbf{The model presented explains the increasing of viscosity, heat
capacity and sound velocity in an aqueous system by enhancing the cooperative
units with ice-like water structure - clustrons and polyclustrons (Table) due
to self-organization in water -macromolecular systems.}

\textbf{The density of ice-like structures in clustrons and polyclustrons is
lower than in usual bulk liquid water.}

\textbf{Consequently, one can predict that the free volume of water will
increase as the result of self-organization in aqueous systems of macromolecules.}

\textbf{Such a type of effect was revealed in our microcalorimetry study of
large-scale and small-scale protein dynamics contributions in the resulting
heat capacity of solvent (K\"{a}iv\"{a}r\"{a}inen et al., 1993). The
additional free volume }$(v_{f})$\textbf{\ in\thinspace\thinspace
[}$0.2-0.3\%]\;(w/v)$\textbf{\ concentration of different proteins is quite
close to the volume occupied by macromolecules in the solution itself. The
}$v_{f}$\textbf{\ is dependent on the active site state (ligand-dependent) and
large-scale dynamics of proteins. The more intensive the large-scale
pulsations of proteins and their flexibility and the less its effective
volume, the smaller the additional free volume solvent. This correlation could
be resulted from acoustic impulses generated by pulsing protein. This
additional acoustic noise can stimulate dissociation (melting) of water
clusters - librational effectons with saturated hydrogen bonds in a system
(K\"{a}iv\"{a}r\"{a}inen et al., 1993). In therms of mesoscopic model it means
increasing the probability of }$[lb/tr]$\textbf{\ convertons
excitation.\medskip\ }

\textbf{The vicinal water in combination with osmotic processes could be
responsible for coordinated intra-cell spatial and dynamic reorganizations.}

\medskip

\begin{center}
{\large 7. Osmosis and solvent activity. Traditional and mesoscopic approach}
\end{center}

\smallskip

It was shown by Van't Hoff in 1887 that osmotic pressure $(\Pi)$ in the dilute
concentration of solute (c) follows a simple expression:%

\begin{equation}
\Pi=\text{ RTc}\tag{24}%
\end{equation}
This formula can be obtained from an \textit{equilibrium condition }between a
solvent and an ideal solution after saturation of diffusion process of the
solvent through a semipermeable membrane:%

\[
\mu_{1}^{0}(P)=\mu_{1}(P+\Pi,X_{i})
\]
where $\mu_{1}^{0}$ and $\mu_{1}$ are the chemical potentials of a pure
solvent and a solvent in solution;$\;P$ - external pressure; $\Pi$ - osmotic
pressure; $X_{1}$ is the solvent fraction in solution.

At equilibrium $d\mu^{0}_{1} = d\mu_{1} = 0$ and%

\begin{equation}
d\mu_{1}=\left[  {\frac{\partial\mu_{1}}{\partial P_{1}}}\right]  _{X_{1}%
}dP_{1}+\left[  {\frac{\partial\mu_{1}}{\partial X_{1}}}\right]  _{P_{1}%
}dX_{1}=0\tag{24a}%
\end{equation}
Because%

\begin{equation}
\mu_{1}=\left(
\begin{array}
[c]{c}%
\partial G/\partial n_{1}%
\end{array}
\right)  _{P,T}=\mu_{1}^{0}+RT\ln X_{1}\tag{25}%
\end{equation}

then%

\begin{equation}
\left(  {\frac{\partial\mu_{1}}{\partial P_{1}}}\right)  _{X_{1}}=\left(
{\frac{\partial^{2}G}{\partial P\partial n_{1}}}\right)  _{P,T,X}=\left(
{\frac{\partial V}{\partial n_{1}}}\right)  =\overline{V}_{1}\tag{26}%
\end{equation}
where $\overline{V}_{1}$ is the partial molar volume of the solvent. For
dilute solution: $\bar{V}_{1}\simeq V_{1}^{0}\;($molar volume of pure solvent).

From (25) we have:%

\begin{equation}
{\frac{\partial\mu_{1}}{\partial X_{1}}}=RT\left(  {\frac{\partial\ln X_{1}%
}{\partial X_{1}}}\right)  _{P,T}\tag{27}%
\end{equation}
Putting (26) and (27) into (24a) we obtain:%

\[
dP_{1}=-{\frac{RT}{V_{1}^{0}X_{1}}}dX_{1}
\]
Integration:%

\begin{equation}
\overset{p+\pi}{\underset{P}{\int}}dP_{1}=-{\frac{RT}{V_{1}^{0}}}%
\overset{x_{1}}{\underset{1}{\int}}d\ln X_{1}\tag{28}%
\end{equation}
gives:%

\begin{equation}
\Pi=-{\frac{RT}{V_{1}^{0}}}\ln X_{1}=-{\frac{RT}{V_{1}^{0}}}\ln(1-X_{2}%
)\tag{29}%
\end{equation}
and for the dilute solution $(X_{2}\ll1)$ we finally obtain Van't Hoff equation:%

\begin{equation}
\Pi={\frac{RT}{V_{1}^{0}}}X_{2}\cong RT{\frac{n_{2}/n_{1}}{V_{1}^{0}}}=\text{
RTc}\tag{30}%
\end{equation}

where%

\begin{equation}
X_{2}=n_{2}/(n_{1}+n_{2})\cong n_{2}/n_{1}\tag{31}%
\end{equation}
and%

\begin{equation}
{\frac{n_{2}/n_{1}}{V_{1}^{0}}}=c\tag{32}%
\end{equation}
Considering a real solution, we only substitute solvent fraction $X_{1}$ in
(29) by solvent activity: $X_{1}\rightarrow a_{1}$. Then taking into account
(25), we can express osmotic pressure as follows:%

\begin{equation}
\Pi=-{\frac{RT}{\bar{V}_{1}}}\ln a_{1}={\frac{\Delta\mu_{1}}{\bar{V}_{1}}%
}\tag{33}%
\end{equation}

where: $\Delta\mu_{1}$ = $\mu_{1}^{0}$ - $\mu_{1}$ is the difference between
the chemical potentials of a pure solvent and the one perturbed by solute at
the starting moment of osmotic process, i.e. the driving force of osmose;
$\bar V_{1}\cong V_{1}$ is the molar volume of solvent at dilute solutions.

Although the osmotic effects are widespread in Nature and are very important,
especially in biology, the physical mechanism of osmose remains unclear
(Watterson, 1992).

The explanation following from Van't Hoff equation (30) and pointing that
osmotic pressure is equal to that induced by solute molecules, if they are
considered as an ideal gas in the same volume at a given temperature is not satisfactory.

\textbf{The osmose phenomenon can be explained quantitatively on the basis of
our mesoscopic theory} \textbf{and state equation (11.7). To this end, we have
to introduce the rules of conservation of the main internal parameters of
solvent in the presence of guest (solute) molecules or particles:}%

\begin{equation}
\left.
\begin{array}
[c]{c}%
1\text{. Internal pressure of solvent: }P_{\text{in}}=\text{ const}\\
2\text{. The total energy of solvent: }U_{\text{tot}}=\text{ const}%
\end{array}
\right\} \tag{34}%
\end{equation}
This conservation rules can be considered as the consequence of Le Chatelier principle.

Using (11.6), we have for the pure solvent and the solvent perturbed by a
solute the following two equations, respectively:%

\begin{equation}
P_{\text{in}}={\frac{U_{\text{tot}}}{V_{fr}^{0}}}\left(  1+{\frac{V}{T_{k}}%
}\right)  -P_{\text{ext}}\tag{35}%
\end{equation}%
\begin{equation}
P_{\text{in}}^{1}={\frac{U_{\text{tot}}^{1}}{V_{fr}^{1}}}\left(
1+{\frac{V_{1}}{T_{k}^{1}}}\right)  -P_{\text{ext}}^{1},\tag{36}%
\end{equation}
where:%

\begin{equation}
V_{fr}^{0}={\frac{V_{0}}{n^{2}}}\text{ \ \ \ and\ \ \ }V_{fr}^{1}={\frac
{V_{0}}{n_{1}^{2}}}\tag{37}%
\end{equation}
are the free volumes of pure solvent and solvent in presence of solute (guest)
molecules as a ratio of molar volume of solvent to correspondent value of
refraction index.

The equilibrium conditions \textit{after osmotic process saturation}, leading
from our conservation rules (34) are%

\begin{align}
P_{\text{in}}  & =P_{\text{in}}\text{ \ \ when \ \ }P_{\text{ext}%
}=P_{\text{ext}}+\Pi\tag{38}\\
U_{\text{tot}}  & =V+T_{k}=V_{1}+T_{k}^{1}=U_{\text{tot}}^{1}\tag{39}%
\end{align}

From (39) we have:%

\begin{equation}
\text{Dif }=T_{k}-T_{k}^{1}=V_{1}-V\tag{40}%
\end{equation}
The index (1) denote perturbed solvent parameters.

Comparing (35) and (36) and taking into account (37 - 39), we obtain a new
formula for osmotic pressure:%

\begin{equation}
\Pi={\frac{n^{2}}{V_{0}}}U_{\text{tot}}\left[  {\frac{n_{1}^{2}T_{k}%
-n^{2}T_{k}^{1}}{T_{k}T_{k}^{1}}}\right] \tag{41}%
\end{equation}

where: $n,\;V_{0},\;U_{\text{tot}}\;$ and $\;T_{k}$ are the refraction index,
molar volume, total energy and total kinetic energy of a pure solvent,
respectively; $T_{k}^{*}$ and $n_{1}$ are the total kinetic energy and
refraction index of the solvent in the presence of guest (solute) molecules,
that can be calculated from the mesoscopic theory (eq.4.33).

For the case of dilute solutions, when\ $T_{k}T_{k}^{1}\cong T_{k}^{2}$
\thinspace and $\;n\cong n_{1},\;$the eq.(41) can be simplified:%

\begin{equation}
\Pi={\frac{n^{2}}{V_{0}}}\left(  {\frac{U_{\text{tot}}}{T_{k}}}\right)
^{2}\left(  T_{k}-T_{k}^{1}\right) \tag{42}%
\end{equation}

or using (40):%

\begin{equation}
\Pi={\frac{n^{2}}{V_{0}}}\left(  {\frac{U_{\text{tot}}}{T_{k}}}\right)
^{2}\left(  V_{1}-V\right) \tag{43}%
\end{equation}

The ratio:%

\begin{equation}
S=T_{k}/U_{\text{tot}}\tag{44}%
\end{equation}
is generally known as a structural factor (see eq. 2.46):

We can see from (42) and (43) that osmotic pressure is proportional to the
difference between total kinetic energy of a free solvent $(T_{k})$ and that
of the solvent perturbed by guest molecules:%

\[
\Delta T_{k}=T_{k}-T_{k}^{1}
\]
or related difference between the total potential energy of perturbed and pure solvent:

$\Delta V=V_{1}-V\;\;$ where: $\Delta T_{k}=\Delta V\equiv$ Dif\ \ (see Fig. 3).

As far $\Delta$T$_{k}>$ 0 \ and\ $\Delta V>0$, it means that:%

\begin{equation}%
\begin{array}
[c]{c}%
T_{k}>T_{k}^{1}\\
\text{or}\\
V_{1}>V
\end{array}
\tag{45}%
\end{equation}
Theoretical temperature dependence of the difference
\[
Dif=\Delta T_{k}=\Delta V
\]
calculated from (42) or (43) at constant osmotic pressure: $\Pi\equiv Pos=8$
atm, pertinent to blood is presented on Fig. 3.

The next Fig. 4 illustrate theoretical temperature dependence of osmotic
pressure (43) in blood at the constant value of \textbf{Dif }$=6.7\cdot
10^{-3}(J/M)$, corresponding on Fig. 3 to physiological temperature $(37^{0})
$.

The ratios of this \textbf{Dif} value to total potential (V) and total kinetic
energy $(T_{k})$ of pure water at $37^{0}($see Fig. 5b) are equal to:%

\begin{align*}
(\text{Dif}/V)  & \simeq{\frac{6.7\cdot10^{-3}}{1.3\cdot10^{4}}}\cong
5\cdot10^{-7}\;\text{ and\ }\\
\text{ (Dif}/T_{k})  & \simeq{\frac{6.7\cdot10^{-3}}{3.5\cdot10^{2}}}%
\cong2\cdot10^{-5}%
\end{align*}
i.e. the relative changes of the solvent potential and kinetic energies are
very small.

\begin{center}%
\begin{center}
\includegraphics[
height=2.2693in,
width=4.9493in
]%
{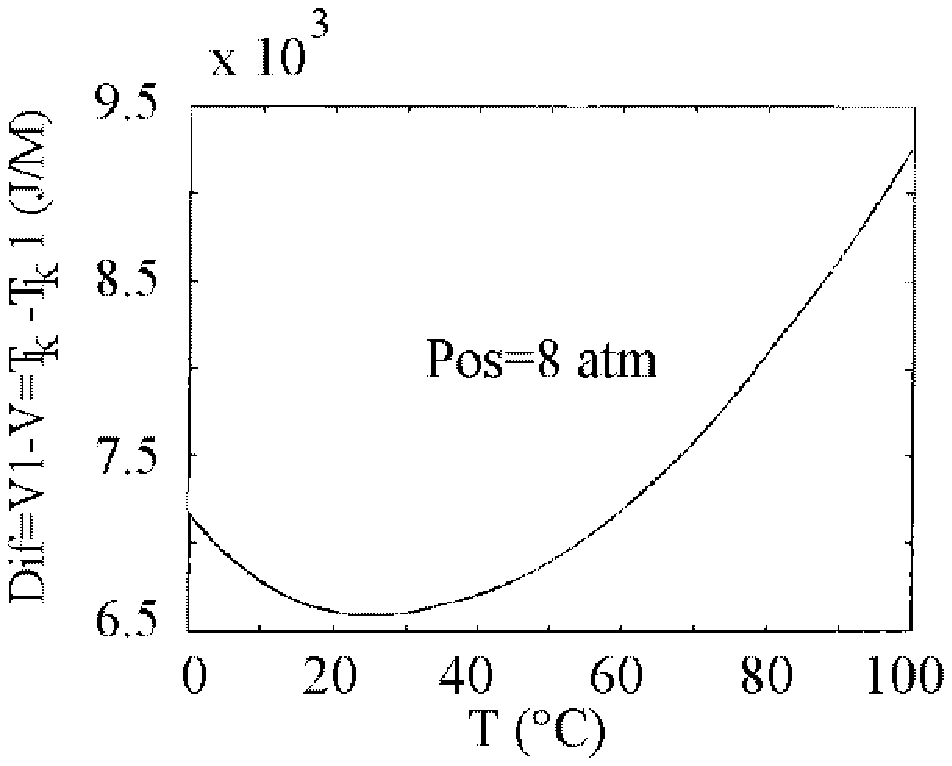}%
\end{center}
\medskip
\end{center}

\begin{quotation}
\textbf{Fig.~3. }Theoretical temperature dependence of the difference:
\textbf{Dif }$=V_{1}-V=T_{k}-T_{k}^{1}$ at constant osmotic pressure:
$\Pi\equiv Pos=8$ atm, characteristic for blood. The computer calculations
were performed using eqs. (42) or (43).
\end{quotation}

\medskip

For each type of \textit{concentrated macromolecular solutions the optimum
amount of water is needed to minimize the potential energy of the system
}determined mainly by clusterphilic interactions. The conservation rules (34)
and self-organization in solutions of macromolecules (clustron formation) may
be responsible for the \textit{driving force of osmos}e in the different
compartments of biological cells.

Comparing (43) and (33) and assuming equality of the molar volumes $V_{0}%
=\bar{V}_{1}$, we find a relation between the difference in potential energies
and chemical potentials $(\Delta\mu)$ of unperturbed solvent and that
perturbed by the solute:%

\begin{equation}
\Delta\mu=\mu_{1}^{0}-\mu_{1}=n^{2}\left(  {\frac{U_{\text{tot}}}{T_{k}}%
}\right)  ^{2}\left(  V_{1}-V\right) \tag{46}%
\end{equation}

\begin{center}%
\begin{center}
\includegraphics[
height=2.0825in,
width=4.9658in
]%
{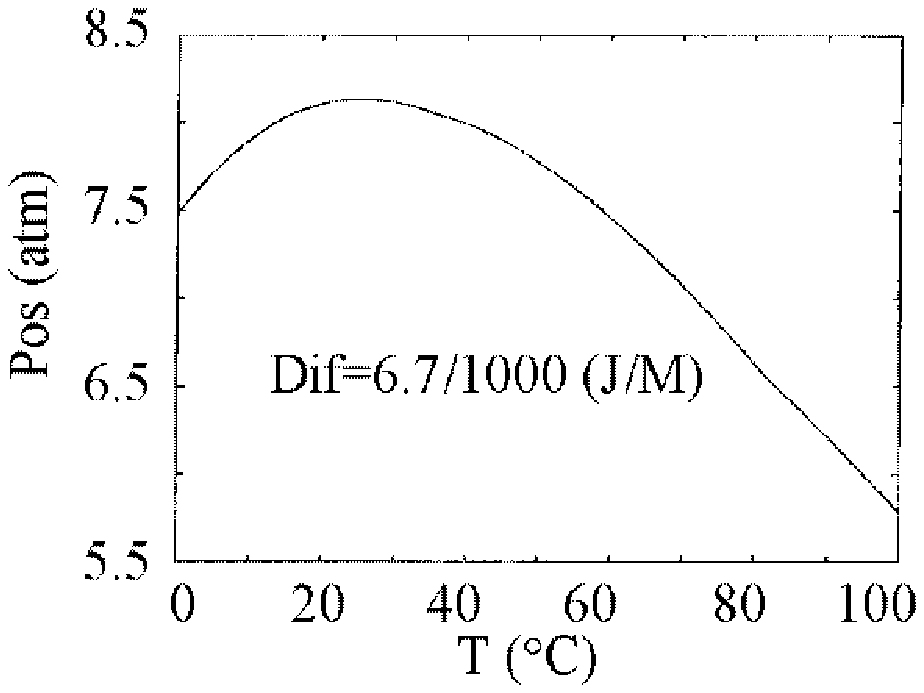}%
\end{center}
\medskip
\end{center}

\begin{quotation}
\textbf{Fig.~4. }Theoretical temperature dependence of osmotic pressure (eq.
43) in blood at constant value of difference: Dif $=\Delta T=\Delta
V=6.7\cdot10^{-3}J/$M. This value in accordance with Fig.3 corresponds to
physiological temperature $(37^{0})$.
\end{quotation}

\smallskip

\textbf{The results obtained above mean that solvent activity }$(a_{1}%
)$\textbf{\ and a lot of other thermodynamic parameters for solutions can be
calculated on the basis of our hierarchic concept:}%

\begin{equation}
a_{1}=\exp\left(  -{\frac{\Delta\mu}{RT}}\right)  =\exp\left[  -\left(
{\frac{n}{S}}\right)  ^{2}{\frac{V_{1}-V}{RT}}\right] \tag{47}%
\end{equation}

where: $S=T_{k}/U_{\text{tot}}$ is a structural factor for the solvent.

The molar coefficient of activity is:%

\begin{equation}
y_{i}=a_{i}/c_{i},\tag{48}%
\end{equation}
where%

\begin{equation}
c_{i}=n_{i}/V\tag{49}%
\end{equation}
is the molar quantity of i-component $(n_{i})$ in of solution (V - solution
volume in liters).

The molar activity of the solvent in solution is related to its vapor pressure
$(P_{i})$ as:%

\begin{equation}
a_{i}=P_{i}/P_{i}^{0}\tag{50}%
\end{equation}
where: P$_{i}^{0}$ is the vapor pressure of the pure solvent. Theoretical
temperature dependence of water activity $(a_{1})$ in blood at constant
difference: Dif $=\Delta T=\Delta V=6.7\cdot10^{-3}J/M$ \thinspace\thinspace
is presented on Fig. 5.

\begin{center}%
\begin{center}
\includegraphics[
height=2.1594in,
width=4.9658in
]%
{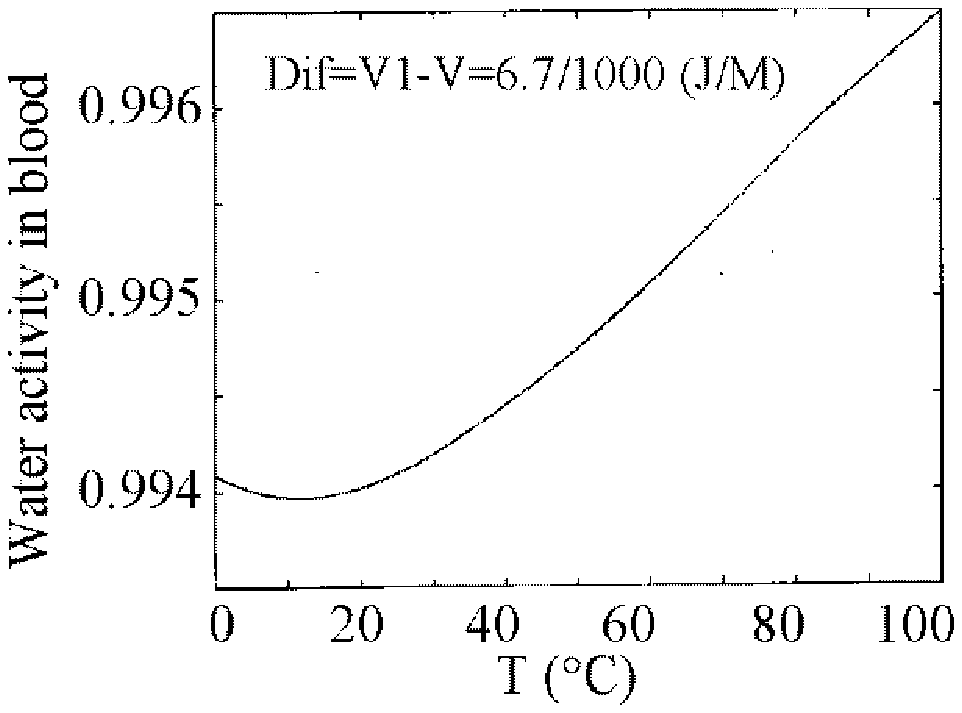}%
\end{center}
\medskip
\end{center}

\begin{quotation}
\textbf{Fig.~5. }Theoretical temperature dependence of water activity
$(a_{1})\;\;(eq.47)$ in blood at constant difference:\ $Dif=\Delta T=\Delta
V=6.7\cdot10^{-3}J/M.$
\end{quotation}

\medskip

Another \textit{colligative parameter }such as low temperature shift of
freezing temperature of the solvent $(\Delta T_{f})$ in the presence of guest
molecules also can be calculated from (47) and (13):%

\begin{equation}
\Delta T_{f}=-{\frac{R(T_{f}^{0})^{2}}{\Delta H}}\ln a_{1}{\frac{(T_{f}%
^{0})^{2}}{\Delta H\cdot T}}\left(  {\frac{n}{S}}\right)  ^{2}(V_{1}%
-V)\tag{51}%
\end{equation}

where: T$_{f}^{0}$ is the freezing temperature of the pure solvent; T is the
temperature corresponding to the conditions of calculations of $V_{1}(T)$ and
V(T) from (eqs. 4.33 and 4.36 of [1,2]).

The partial molar enthalpy $(\bar H_{1})$ of solvent in solution are related
to solvent activity like:%

\begin{equation}
\overline{H}_{1}=\overline{H}_{1}^{0}-RT^{2}{\frac{\partial\ln a_{1}}{\partial
T}}=\overline{H}_{1}^{0}+\overline{L}_{1}^{0}\tag{52}%
\end{equation}

where $\overline{H}_{1}^{0}$ is the partial enthalpy of the solvent at
infinitive dilution;%

\begin{equation}
\bar{L}_{1}=-RT^{2}{\frac{\partial\ln a_{1}}{\partial T}}=T^{2}{\frac
{\partial}{\partial T}}\left[  \left(  {\frac{n}{S}}\right)  ^{2}{\frac
{V_{1}-V}{T}}\right] \tag{53}%
\end{equation}

is the relative partial \textbf{molar enthalpy of solvent }in a given solution.

From (52) we obtain partial molar heat capacity as:%

\begin{equation}
C_{p}^{1}={\frac{\partial}{\partial T}}(H_{1})=C_{p}^{0}-R\left(  T^{2}%
{\frac{\partial^{2}\ln a_{1}}{\partial T^{2}}}+2T{\frac{\partial\ln a_{1}%
}{\partial T}}\right) \tag{54}%
\end{equation}
An analogous equation exists for the \textit{solute }of this solution as well
as for partial \textit{molar volume }and other important parameters of the
solvent, \textit{including solvent activity }(Godnev et al., 1982).

\textbf{It is obvious, the application of Hierarchic theory to solvent
activity determination might be of great practical importance for different
processes in chemical and colloid technology. Our theory based Comprehensive
Analyzer of mattr Properties (CAMP - see www.karelia.ru/ \symbol{126}alexk)
may be conveniant tool for monitoring of such processes.}

\bigskip

\subsubsection{{\protect\LARGE \ \noindent References}}

\smallskip

\smallskip

\begin{quotation}
\textbf{Aksnes G., Asaad A.N. Influence of the water structure on chemical
reactions in water. A study of proton-catalyzed acetal hydrolysis. Acta Chem.
Scand. }$1989,43,726-734$\textbf{.}

\textbf{Aksnes G., Libnau O. Temperature dependence of esther hydrolysis in
water. Acta Chem.Scand. }$1991,45,463-467$\textbf{.}

\textbf{Aliotta F., Fontanella M.E., Magazu S. Sound propagation in
thyxotropic strucures. Phys. Chem. Liq. 1990,}

\textbf{Benassi P., D'Arrigo G., Nardone M. Brilouin light scattering in low
temperature water-ethanol solutions. J.Chem.Phys. }$1988,89,4469-4477$\textbf{.}

\textbf{Bertolini D., Cassetari M., Grigolini P., Salvetti G. and Tani A. The
mesoscopic systems of water and other complex systems.
J.Mol.Liquids,\thinspace\thinspace\thinspace\thinspace}$1989,\;41,\,251$\textbf{.}

\textbf{Bertolini D., Cassetari M., Salvetti G., Tombari E., Veronesi S.,
Squadrito G. Il nuovo Cim. }$1992,\,14D,\;199$\textbf{.}

\textbf{Cantor C.R., Schimmel P.R. Biophysical Chemistry. W.H.Freemen and
Company, San Francisco, 1980.}

\textbf{Clegg J. S. On the physical properties and potential roles of
intracellular water. Proc.NATO Adv.Res.Work Shop. 1985.}

\textbf{Clegg J.S. and Drost-Hansen W. On the biochemistry and cell physiology
of water. In: Hochachka and Mommsen (eds.). Biochemistry and molecular biology
of fishes. Elsevier Science Publ. vol.1, Ch.1, pp.1-23, 1991.}

\textbf{D'Aprano A., Donato Ines., Liveri V.T. Molecular association of n-
alcohols in nonpolar solvents. Excess volumes and viscosities of n-
pentanol+n-octane mixtures at 0, 5, 25, 35 and }$45^{0}$\textbf{C.
J.Solut.Chem. }$1990a,\mathbf{1}\mathbf{9}\mathbf{,}711-720$\textbf{.}

\textbf{D'.Aprano A., Donato I., Liveri V.T. Molecular interactions in 1-
pentanol + 2-butanol mixtures: static dielectric constant, viscosity and
refractive index investigations at 5, 25, and }$45^{0}$\textbf{C.
J.Solut.Chem. }$1990b,18,785-793$\textbf{.}

\textbf{D'Aprano A. and Donato I. Dielectric polarization and Polarizability
of 1-pentanol + n-octane mixtures from static dielectric constant and
refractive index data at 0, 25 and }$45^{0}$\textbf{. J.Solut.Chem.
}$1990c,\,19,\,883-892$\textbf{.}

\textbf{D'Arrigo G., Paparelli A. Sound propagation in water-ethanol mixtures
at low temperatures. I.Ultrasonic velocity. J.Chem.Phys. }$1988a,\,88$%
\textbf{, No.}$1,405-415$\textbf{.}

\textbf{D'Arrigo G., Paparelli A. Sound propagation in water-ethanol mixtures
at low temperatures. II.Dynamical properties. J.Chem.Phys. }$1988b,\,88$%
\textbf{, No.}$12,7687-7697$\textbf{.}

\textbf{D'Arrigo G., Paparelli A. Anomalous ultrasonic absorption in
alkoxyethanls aqeous solutions near their critical and melting points.
J.Chem.Phys. 1989, 91, No.}$4,\,2587-2593$\textbf{.}

\textbf{D'Arrigo G., Texiera J. Small-angle neutron scattering study of
}$D_{2}O $\textbf{-alcohol solutions. J.Chem.Faraday Trans. }%
$1990,\,86,\,1503-1509$\textbf{.}

\textbf{Del Giudice E., Dogulia S., Milani M. and Vitello G. A quantum field
theoretical approach to the collective behaviour of biological systems.
Nuclear Physics\ }$1985,$\textbf{\ }$B251[FS13],375-400$\textbf{.}

\textbf{Del Guidice E. Doglia S., Milani M. Spontaneous symmetry breaking and
electromagnetic interactions in biological systems. Physica Scripta.
}$1988,38,505-507$\textbf{.}

\textbf{Drost-Hansen W. In: Colloid and Interface Science. Ed. Kerker M.
Academic Press, New York, 1976, p.267.}

\textbf{Drost-Hansen W., Singleton J. Lin. Our aqueous heritage: evidence for
vicinal water in cells. In: Fundamentals of Medical Cell Biology, v.3A,
Chemisrty of the living cell, JAI\ Press Inc.,1992, p.157-180.}

\textbf{Etzler F.M., Conners J.J. Structural transitions in vicinal water:
pore size and temperature dependence of the heat capacity of water in small
pores. Langmuir 1991, 7, 2293-2297.}

\textbf{Etzler F.M., White P.J. The heat capacity of water in silica pores. J.
Colloid and Interface Sci., 1987, 120, 94-99.}

\textbf{Farsaci F., Fontanella M.E., Salvato G., Wanderlingh F. and Giordano
R., Wanderlingh U. Dynamical behaviour of structured macromolecular solutions.
Phys.Chem. Liq. }$1989,\,20,\,205-220$\textbf{.}

\textbf{Fontaine A. et al., Phys Rev. Lett. }$1978,\,\mathbf{4}\mathbf{1}%
\mathbf{,}504$\textbf{.}

\textbf{Fr\"{o}hlich H. Phys.Lett.\ 51\ (1975)\ 21.}

\textbf{Fr\"{o}hlich H. Proc. Nat. Acad. Sci. USA \ }$1975,\,72,\,4211$\textbf{.}

\textbf{Giordano R., Fontana M.P., Wanderlingh F. J.Chem.Phys. }%
$\,1981a,\,74,\,2011$\textbf{.}

\textbf{Giordano R. et al. Phys.Rev. }$\,1983b,\,A28,\,3581$\textbf{.}

\textbf{Giordano R., Salvato G., Wanderlingh F., Wanderlingh U. Quasielastic
and inelastic neutron scattering in macromolecular solutions.
Phys.Rev.A.\thinspace}$\,\,1990,41,689-696$\textbf{.}

\textbf{Glansdorf P., Prigogine I. Thermodynamic theory of structure,
stability and fluctuations. Wiley and Sons, N.Y., 1971.}

\textbf{Gordeyev G.P., Khaidarov T. In: Water in biological systems and their
components. Leningrad University Press, Leningrad, 1983, p.3 (in Russian).}

\textbf{Haken H. Information and selforganization. Springer, Berlin, 1988.}

\textbf{Haken H. Synergetics, computers and cognition. Springer, Berlin, 1990.}

\textbf{Ise N. and Okubo T. Accounts of Chem. Res. }$1980,13,303$\textbf{.}

\textbf{Ise N., Matsuoka H., Ito K., Yoshida H. Inhomogenity of solute
distribution in ionic systems. Faraday Discuss. Chem. Soc. }$1990,\,\mathbf{9}%
\mathbf{0}\mathbf{,\,}153-162$\textbf{.}

\textbf{Ito K., Yoshida H., Ise N. Void Structure in colloid Dispersion.
Science, }$\,1994,\,263,\,66$\textbf{.}

\textbf{K\"{a}iv\"{a}r\"{a}inen A.I. Solvent-dependent flexibility of proteins
and principles of their function. D.Reidel Publ.Co., Dordrecht, Boston,
Lancaster, 1985,\thinspace pp.290.}

\textbf{K\"{a}iv\"{a}r\"{a}inen A.I. The noncontact interaction between
macromolecules revealed by modified spin-label method. Biofizika
(USSR}$)\;1987,\,32,\,536$\textbf{.}

\textbf{K\"{a}iv\"{a}r\"{a}inen A.I. Thermodynamic analysis of the system:
water-ions-macromolecules. Biofizika (USSR}$),\,1988,\,33,\,549$\textbf{.}

\textbf{K\"{a}iv\"{a}r\"{a}inen A.I. Theory of condensed state as a
hierarchical system of quasiparticles formed by phonons and three-dimensional
de Broglie waves of molecules. Application of theory to thermodynamics of
water and ice. J.Mol.Liq. }$1989a,\,41,\,53-60$\textbf{.}

\textbf{K\"{a}iv\"{a}r\"{a}inen A.I. Mesoscopic theory of matter and its
interaction with light. Principles of selforganization in ice, water and
biosystems. University of Turku, Finland\thinspace1992, pp.275.}

\textbf{K\"{a}iv\"{a}r\"{a}inen A., Fradkova L., Korpela T. Separate
contributions of large- and small-scale dynamics to the heat capacity of
proteins. A new viscosity approach. Acta Chem.Scand. }$1993,47,456-460$\textbf{.}

\textbf{Lagrage P., Fontaine A., Raoux D., Sadoc A., Miglardo P.
\thinspace\thinspace J.Chem. Phys. }$\,1980,\,\mathbf{7}\mathbf{2}%
\mathbf{,}3061$\textbf{.}

\textbf{Lumry R. and Gregory R.B. Free-energy managment in protein reactions:
concepts, complications and compensations. In book: The fluctuating enzyme. A
Wiley-Interscience publication. 1986, p. 341- 368.}

\textbf{Nemethy G., Scheraga H.A. J.Chem.Phys. \thinspace}%
$1962\mathbf{,\;36,\,}3382$\textbf{.}

\textbf{Peschel G. Adlfinger K.H. J. Coll. Interface Sci., 1970, v.34 (4), p.505.}

\textbf{Sadoc A., Fountaine A., Lagarde D., Raoux D. J.Am.Chem.Soc.\thinspace
\thinspace}$1981,$\textbf{\ }$\mathbf{1}\mathbf{0}\mathbf{3},6287$\textbf{.}

\textbf{Tereshkevitch M.O., Kuprin F.V., Kuratova T.S., Ivishina G.A. J. Phys.
Chem. (USSR}$)\;1974,\,48,\,2498$\textbf{.}

\textbf{Watterson J. Solvent cluster size and colligative properties. Phys.
Chem.Liq. }$1987,16,317-320$\textbf{.}

\textbf{Watterson J. The role of water in cell architecture. Mol.Cell.Biochem.
}$1988\mathbf{,}\mathbf{7}\mathbf{9}\mathbf{,}101-105$\textbf{.}

\textbf{Yoshida H., Ito K., and Ise N. Colloidal crystal growth. J. Chem. Soc.
Faraday Trans., }$1991,87(3),371-378$\textbf{.}
\end{quotation}
\end{document}